\documentclass{article}
\usepackage[utf8]{inputenc}
\usepackage{graphicx}
\usepackage[a4paper, total={6.5in, 9in}]{geometry}
\usepackage{cite}
\usepackage[english]{babel}
\usepackage{lmodern}
\usepackage[T1]{fontenc}
\usepackage{amsmath,amsfonts,amsthm}
\usepackage{booktabs}
\usepackage{siunitx}
\usepackage[labelfont=bf]{caption}
\usepackage{float}
\usepackage{subcaption}
\usepackage{braket}
\usepackage[utf8]{inputenc}
\usepackage{siunitx}
\usepackage[symbol]{footmisc}
\usepackage{authblk}
\usepackage{color}
\usepackage{hyperref}
\hypersetup{colorlinks=false,linkcolor=black,filecolor=black,urlcolor=black,citecolor=black}
\usepackage{lineno}

\usepackage{lipsum}

\title{\textbf{Collective cell mechanics of small-organoid morphologies}}
\author[1,2,*]{Jan Rozman}
\author[1,3]{Matej Krajnc}
\author[1,2]{Primo\v z Ziherl}

\affil[1]{ Jo\v zef Stefan Institute, Jamova 39, SI-1000 Ljubljana, Slovenia}
\affil[2]{Faculty of Mathematics and Physics, University of Ljubljana, Jadranska 19, SI-1000 Ljubljana, Slovenia}
\affil[3]{Lewis--Sigler Institute for Integrative Genomics, Princeton University, Washington Road, Princeton NJ 08540, USA}
\affil[*]{corresponding author: jan.rozman@ijs.si}

\begin{document}

\maketitle

\begin{abstract}
The study of organoids, artificially grown cell aggregates with the functionality and small-scale anatomy of real organs, is one of the most active areas of research in biology and biophysics, yet the basic physical origins of their different morphologies remain poorly understood. Here we propose a mechanistic theory of small-organoid morphologies. Using a 3D surface-tension-based vertex model, we reproduce the characteristic shapes, ranging from branched and budded structures to invaginated shapes. We find that the formation of branched morphologies relies strongly on junctional activity, enabling temporary aggregations of topological defects in cell packing. To elucidate our numerical results, we develop an effective elasticity theory, which allows one to estimate the apico-basal polarity from the organoid-scale modulation of cell height. Our work provides a generic interpretation of the observed small-organoid morphologies, highlighting the role of physical factors such as the differential surface tension, cell rearrangements, and tissue growth.
\end{abstract}

\section*{Introduction}

In less than a decade, organoids became one of the most interesting topics in cells and tissue biology, organogenesis, developmental biology, and study of disease~\cite{Lancaster14,Clevers16}. Organoids are aggregates of cells grown {\sl in vitro} so as to form miniature replicas of a given organ such as intestine, kidney, lung, and brain~\cite{Chen17,Karzbrun18}. In most cases, they consist of a closed shell of sheet-like tissue enclosing  a lumen, and they have the same microscopic morphology as the mimicked organ itself, e.g., the villus-crypt pattern of the mammalian intestine. Organoids are often grown from embryonic stem cells driven so as to develop a given tissue identity and then cultured in a medium such as Matrigel. They can also be grown from adult tissue- or stem cells~\cite{Huch13,Sato09}.

One of the defining features of organoids is their physical form. Their initially simple shape transforms into a given morphology as cells divide and mature~\cite{Buske12}. The most common organoid morphologies include budded and branched shapes with various spherical or finger-like protrusions, respectively; some branched shapes develop bifurcating networks. The shape of organoids depends on many factors and processes including the intrinsic physical features of individual cells, cell growth and division rates, cell differentiation, etc. So far, theoretical studies of the role of these factors were mostly carried out by representing the tissue as a continuous concentration field in a model of Cahn--Hilliard type~\cite{Wise08,Yan18} or as an ensemble of spherical entities using discrete models~\cite{Buske11,Buske12}. These approaches account for the biochemical regulation of organoid growth, yielding invaluable insight into, e.g., potential strategies of anticancer therapy~\cite{Yan18}. Recently, a 3D vertex model was employed to study how chemical patterning controlling the local cell growth rate may feed back to the mechanics to determine organoid morphology~\cite{Okuda18a}. Together with results addressing morphogenesis of optic cup organoids~\cite{Okuda18b} and transformations of epithelial shells~\cite{Misra16}, these insights demonstrate that {many features of the observed organoid shapes can be interpreted using simple physical models based on mechanisms arising, naturally, from the underlying cell- and tissue-level biological processes.

Here we enhance such a perspective by analyzing a surface-tension-based vertex model of small-organoid shapes which evolve from a spherical shell of cells. To emphasize the collective mechanical effects, we study model organoids consisting of cells of identical properties; in this respect, our organoids resemble tumor and other spheroids where cell differentiation is often absent.} The diversity of the obtained shapes is striking, since their formation relies exclusively on the most generic cell-scale mechanics such as the apico-basal differential surface tension. %
Our model reproduces almost all experimentally observed morphologies and is further interpreted in terms of a theory of epithelial elasticity, which contains several interesting elements such as the curvature-thickness coupling. Furthermore, we explore the formation of in-plane cell arrangements and find that topological defects in cell packing induced by active rearrangements can act as seeds for branching morphogenesis, leading to out-of-equilibrium branched shapes. We also study the formation of organoid shape during tissue growth through successive cell divisions. Overall, these results reveal generic mechanisms of the formation of shape in small organoids of identical cells.

\section*{Results}

\begin{figure}[h!]
\centerline{\includegraphics{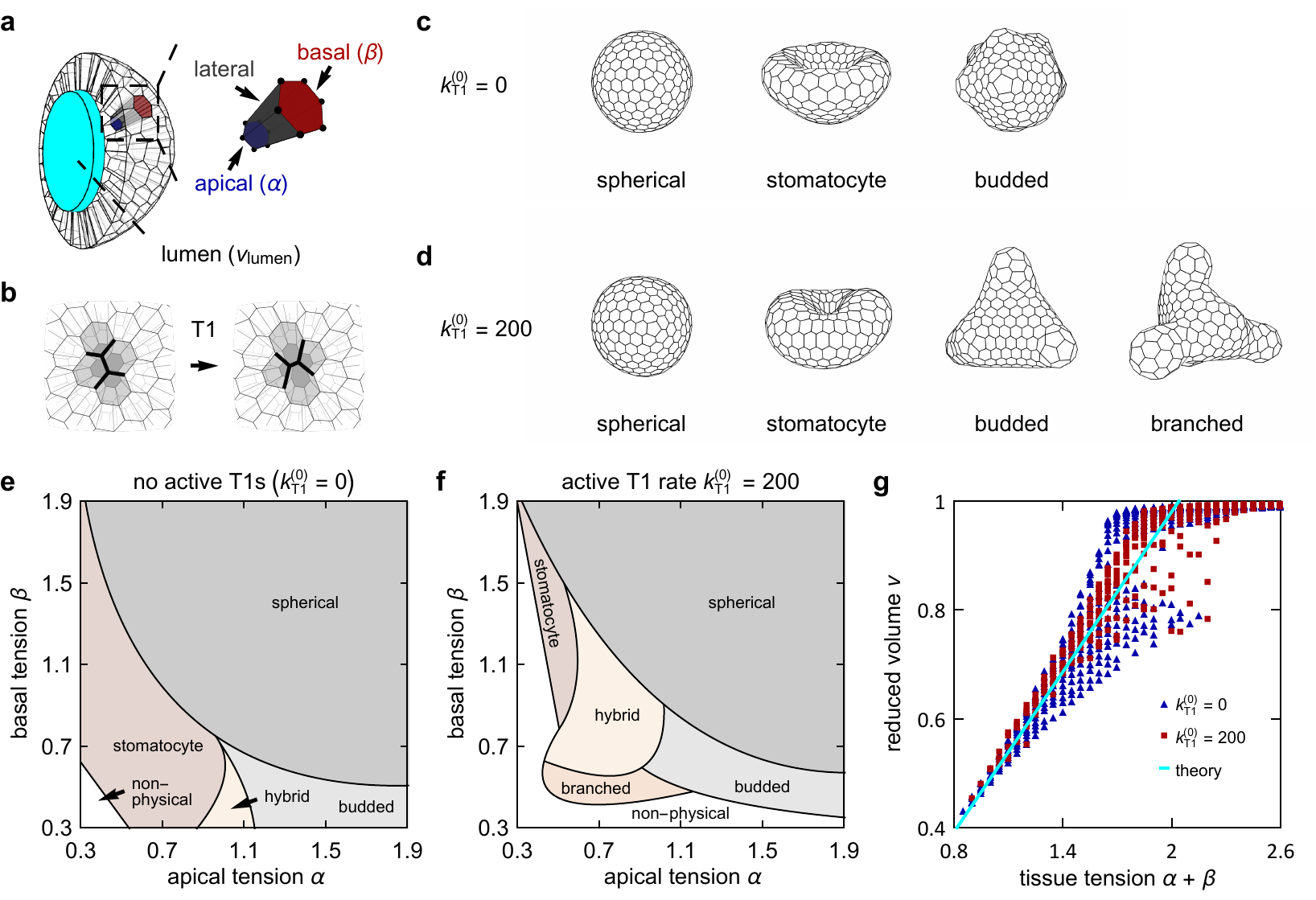}}
\caption{\label{F1} \textbf{Mechanical model reproduces characteristic small-organoid morphologies.} {\bf a}~Schematics of a model organoid with a lumen of volume $v_{\rm lumen}$ (expressed in units of cell volume $V_{\rm cell}$) indicated in cyan, and a cell with apical, lateral, and basal sides; also indicated are the dimensionless apical and basal tensions $\alpha$ and $\beta$, respectively. {\bf b}~T1 transition. {\bf c}~Representative organoid shapes with $N_c=300$ cells, $v_{\rm lumen}=100$, and $k_{\rm T1}^{(0)}=0$:  Spherical $(\alpha=1.2,\>\beta=1.2)$, stomatocyte $(\alpha=0.7,\>\beta=0.5)$, and budded $(\alpha=1.5,\>\beta=0.3)$ morphologies. Panel {\bf d} shows model organoids at the same $N_c$ and $v_{\rm lumen}$ but with $k_{\rm T1}^{(0)}=200$: Spherical $(\alpha=1.2,\>\beta=1.2)$, stomatocyte $(\alpha=0.5,\>\beta=1.1)$, budded $(\alpha=1.1,\>\beta=0.5)$, and branched  ($\alpha=0.7$, $\beta=0.5$) morphologies. {\bf e}~Phase diagram of the $N_c=300$, $v_{\rm lumen}=100$, and {$k_{T1}^{(0)}=0$} shapes in the $(\alpha,\beta)$ plane.  {\bf f}~Phase diagram at same $N_c$ and $v_{\rm lumen}$ as in panel e but with $k_{\rm T1}^{(0)}=200$. {\bf g}~Reduced volume $v$ vs.~tissue tension $\alpha+\beta$. The solid line is a theoretical prediction of the relation in  non-spherical shapes [Eq.~(\ref{eq:redvol_semianalytical})].}
\end{figure}

We model organoids as single-cell-thick epithelial shells enclosing an incompressible fluid-like interior referred to as lumen~(Fig.~\ref{F1}a). The cells too are assumed to be incompressible, and they carry a surface energy with three distinct tensions at the inner apical side, the outer basal side, and the lateral sides where they adhere to their neighbors~\cite{Storgel16}. By considering the surface energy alone, we disregard several elements of cell mechanics such as the acto-myosin cable at the apical surface. At the same time, this simplification allows us to work with fewer model parameters which makes the analysis more accessible. We further assume that all cells are identical in terms of their surface tensions and cell volumes~$V_{\rm cell}$. This allows us to explore a more generic scenario where instead of relying on cell differentiation, complex organoid morphologies arise directly from an interplay between the preferred shape of individual cells and long-range interactions due to lumen incompressibility.

We use a 3D surface-tension-based vertex model~\cite{Krajnc18}, where cells are represented by polyhedra with polygonal apical and basal sides and rectangular lateral sides~(Fig.~\ref{F1}a);  the topologies of the apical and the basal cell networks are identical. Cell shapes are parametrized by the vertex positions $\boldsymbol r_j=(x_j,y_j,z_j)$, where $x_j,y_j,$ and $z_j$ are dimensionless coordinates expressed in units of $V_{\rm cell}^{1/3}$. The dimensionless energy of the organoid, expressed in units of $\Gamma_lV_{\rm cell}^{2/3}$ (where $\Gamma_l$ is the surface tension at the lateral sides) is given by a sum over all $N_c$ cells:  
\begin{equation}
w = \sum_{i=1}^{N_c} \left[\alpha a_a^{(i)}+\beta a_b^{(i)}+\frac{1}{2}a_l^{(i)}\right].
\label{eq:energy1}
\end{equation} 
Here $\alpha$ and $\beta$ are the dimensionless tensions of the apical and basal sides, respectively, expressed in units of $\Gamma_l$~(Fig.~\ref{F1}a), whereas $a_a^{(i)}, a_b^{(i)},$ and $a_l^{(i)}$ are the areas of the apical, basal, and lateral sides of cell $i$, respectively, expressed in units of~$V_{\rm cell}^{2/3}$.

In our model, tissue dynamics includes two concurrent processes: (i) deterministic relaxation of the system in the direction of the energy gradient and (ii) cell rearrangements driven by active junctional noise, which allow our model tissue to fluidize. During each time step, the vertices are first moved according to the overdamped equation of motion ${\rm d}{\boldsymbol r}_j/{\rm d}t=-\nabla_j w$; 
here $\nabla_j$ is the gradient with respect to the dimensionless ${\boldsymbol r}_j$, $w$ is the dimensionless energy, and dimensionless time $t$ is measured in units of $\tau_0=(\Gamma_l\mu)^{-1}$ with $\mu$ being vertex mobility. Additionally, cells are allowed to rearrange through T1~transitions~(Fig.~\ref{F1}b and Methods) following the model from~Ref.~\cite{Krajnc18}. {In particular, active noise at cell-cell junctions due to the stochastic turnover dynamics of molecular motors provides energy fluctuations which can drive the system from one metastable state to another via T1 transitions. These transitions are implemented by swapping the four cells arranged around a given lateral side, which is the physical cell-cell junction.

It is convenient to view the junction projected onto the midplane between the apical and the basal surface---and to treat it as an edge of length halfway between those of the corresponding edges on the apical and the basal surface. Since the thus-redefined cell-cell junctions that have a longer length are associated with a higher energy barrier between the two metastable states, they are less likely to undergo the T1 transition.} To describe this, we employ the previously proposed threshold-based scheme where the probability of a T1~transition in junctions shorter than the threshold $\delta l$ is unity, whereas junctions longer than the threshold undergo the transition with a probability $k_{\rm T1} \delta t / \mathcal{E}$~\cite{Krajnc18}. Here $k_{\rm T1}$ is the rate of active T1 transitions measured in units of $\tau_0^{-1}$, $\delta t = 10^{-4}$ is the time step, and $\mathcal{E}$ is the total number of cell-cell junctions. {We choose $\delta l = 0.15$ (defined as a dimensionless quantity expressed in units of $V_{\rm cell}^{1/3}$) which is between a quarter and a third of the average junction length; we stress that the results do not depend significantly on this choice~({Supplementary Fig.~4b}).} As shown previously, tissues that are subject to this type of active noise behave like viscous fluids with activity-dependent stress-relaxation time scale and the associated effective viscosity~\cite{Krajnc18}.
Therefore, an important aspect of the organoid shapes could be their sensitivity to the changes of the active T1 rate~$k_{\rm T1}$. 

In most cases discussed here, the model parameters include only a handful of dimensionless quantities: The apical and basal surface tensions $\alpha$ and $\beta$, the volume of the lumen relative to the cell volume $v_{\rm lumen}~=~V_{\rm lumen}/V_{\rm cell}$, the total cell number $N_c$, and at most two parameters related to junctional activity. 

\subsection*{\label{sec:pd}Phase diagram}
{
We start by studying the shapes of organoids with fixed cell number $N_c=300$ and lumen volume $v_{\rm lumen}=~100$ and we first consider the case with no active T1 transitions; this limit is denoted by $k_{\rm T1}^{(0)} = 0$ for consistency of notation, with the meaning of $k_{\rm T1}^{(0)}$ explained in the next paragraph. The initially spherical shape evolves into three characteristic types of morphologies depending on the values of $\alpha$ and $\beta$: stomatocyte (cup-like), budded, and spherical~(Fig.~\ref{F1}c and {Supplementary Movie~1}). In the phase diagram in the $(\alpha,\beta)$ plane (Fig.~\ref{F1}e and {Supplementary Fig.~1a}), spherical shapes occupy the region where the tissue tension defined by $\alpha+\beta$ is larger than about 1.9, budded shapes emerge at $\alpha$ larger than about 1 and $\beta$ smaller than about 0.6, whereas stomatocytes occur in the region where $\alpha+\beta$ is smaller than 1.9 but larger than around 0.9 and $\alpha$ is less than about 1. In many shapes, the characteristic features are well-developed so that the classification is unambiguous, but this is not always the case. For example, a generally spherical model organoid with small buds may be regarded either as a spherical or as a budded shape. Shapes found at the boundary between the stomatocyte and the budded-organoid domain contain features of both morphologies and are therefore best referred to as hybrid ({Supplementary Fig.~1c}). Finally, at tissue tensions $\alpha+\beta$ that are smaller than about 0.9 we find model organoids containing one or more cells of anomalous, non-physical shape, or else the organoids start to self-overlap.

We next analyze the case where relaxation is accompanied by an initially high rate of active cell rearrangements with $k_{\rm T1}(t=0)=k_{\rm T1}^{(0)}=200$, which decreases in time to zero as $k_{\rm T1}(t) = k_{\rm T1}^{(0)} (t_{\rm max} - t) / t_{\rm max}$; here $t_{\rm max} = 1000$ is the total simulation time. The finite duration of junctional activity mimics the natural course of morphogenetic events, which, guided by biochemical signals, begin and end at a predefined time; the linear temporal profile is just a simple example of a gradual decrease of activity. In Section "Active cell rearrangements" we show that this choice does not significantly affect the results. Like in the case with no activity, we obtain the budded, stomatocyte, and spherical shapes. However, the active T1 transitions also give rise to the branched morphologies appearing at the apical and basal tensions somewhat smaller than unity, that is in part of  the region occupied by stomatocytes if the junctional activity is absent (Figs.~\ref{F1}d and f, {Supplementary Figs.~1b and~2}, and {Supplementary Movie~2}). Like before, the phase diagram also contains hybrid shapes that feature either multiple invaginations, both invaginations and evaginations, or are devoid of any clear features ({Supplementary Fig.~1d}); these shapes appear at $\alpha\approx\beta$ and tissue tension $\alpha+\beta$ between about 1 and 2. The domain of non-physical shapes is somewhat larger than in absence of active T1 transitions.

The results obtained using the $N_c=300$ organoids are quite representative of our model. This is demonstrated by the four sets of shapes in {Supplementary Fig.~3} which show that with an appropriate rescaling of the preferred lumen volume $v_{\rm lumen}$~(Methods), the type of organoid morphology does not change significantly with the cell number $N_c$ even if $N_c$ is halved or doubled. On the other hand, the fixed lumen volume constraint does play an important role in the model morphologies. This constraint is implemented using an auxiliary harmonic volumetric energy term, which ensures that the actual lumen volume does not depart from the preferred value. If the modulus $K_{\rm lumen}$ of the auxiliary term is small enough compared to the typical values of $\alpha$ and $\beta$ (for example, 0.001) then the actual and the preferred volumes differ considerably and the model organoids lack clear features and are more or less round ({Supplementary Fig.~4a}).}

\subsection*{\label{sec:shape_origin}Origin of shape}
Aside from the role of active T1 transitions discussed in Section ``Active cell rearrangements'', the main mechanisms responsible for the formation of the distinct morphologies are (i) the incompatibility of the preferred surface area of the organoid and the enclosed lumen volume and (ii) the apico-basal differential tension. In particular, the preferred cell width-to-height ratio scales with the tissue tension $\alpha+\beta$ as~${\sim(\alpha+\beta)^{-1}}$ so that small values of $\alpha+\beta$ drive the columnar-to-squamous transition~\cite{Krajnc18}. In turn, this increases the organoid surface area at a fixed lumen volume, causing the formation of complex non-spherical morphologies. A similar mechanism was previously studied within a 2D model of tissue cross section, yielding shapes resembling our 3D small-organoid morphologies~\cite{HocevarBrezavscek12}. 

The area-volume incompatibility can be conveniently quantified by the reduced volume defined by
\begin{equation}
    \label{eq:redvol}
    v=\frac{6\sqrt{\pi}V}{A^{3/2}}\>.
\end{equation}
Here $A$ is the area of the midplane surface located halfway between the basal and the apical side of the organoid and $V$ is the corresponding enclosed volume; $v=1$ for a sphere, and its value decreases with  increasing area at constant volume. We compute the reduced volumes of all 1328 shapes used to construct the phase diagrams in Figs.~\ref{F1}e and f and find that they depend strongly on the tissue tension $\alpha+\beta$~(Fig.~\ref{F1}g and {Supplementary Fig.~5a}). The numerically obtained dependence agrees well with the analytical prediction where we estimate the organoid midplane surface area by considering a flat epithelium of identical cells with regular hexagonal basal and apical sides~\cite{Krajnc18}. {From the force balance along cell height $h$, $\partial w/\partial h=0$, and the relation between (dimensionless) cell volume 1, height $h$, and area $a$, which reads  $1=ha$, we can calculate the equilibrium cell height $h_0=\left (2^{1/3}/3^{1/6}\right)(\alpha+\beta)^{2/3}$ and midplane area $a_0=\left (3^{1/6}/2^{1/3}\right )(\alpha+\beta)^{-2/3}$. 
Inserting the expression for $a_0$ into Eq.~(\ref{eq:redvol}) yields
\begin{equation}
    \label{eq:redvol_semianalytical}
    v=\frac{2^{3/2}3^{3/4}\sqrt{\pi}v_{\rm midplane}}{N_c^{3/2}}\left(\alpha+\beta\right),
\end{equation}
where the volume enclosed by the midplane $v_{\rm midplane}=223$ (expressed in units of $V_c$) is the average  taken from the simulated shapes. The relation between $v$ and $\alpha+\beta$ given by Eq.~(\ref{eq:redvol_semianalytical}) is plotted by the solid cyan line in Fig.~{1}g.
}

While the apico-basal differential tension $\alpha-\beta$ has a rather limited effect on the reduced volume $v$ as witnessed by the limited spread of points in Fig.~\ref{F1}g, we find that it is often crucial for the formation of budded and stomatocyte morphologies. This is because the buds consist of cells with the apical sides smaller than the basal sides whereas invaginations in the stomatocyte shapes require cells where the apical sides are larger than the basal ones. The former cell type is energetically preferred at $\alpha-\beta > 0$ whereas the latter is favored at $\alpha-\beta < 0$. 

{An insightful way of interpreting our small-organoid shapes builds on the analogy with lipid vesicles characterized by spontaneous curvature~\cite{Seifert91}. In particular, (i)~both organoids and vesicles are closed shells with  a well-defined surface area and enclosed volume as well as fluid-like in-plane order in the case of organoids with junctional activity and (ii)~the apico-basal polarity due to the differential tension $\alpha-\beta$ in tissues leads to a preferred conical shape of cells~\cite{Krajnc15,Storgel16} so that the tissue has a certain spontaneous curvature reminiscent of that in asymmetric lipid membranes, e.g., due to inclusions. Although organoids are thick rather than thin shells like vesicles, this comparison is quite illuminating because of both similarities and differences, especially within the context of the theory of elasticity developed in the penultimate section of this paper. As far as the characteristic morphologies are concered, we note that the phase diagrams of organoids and vesicles feature stomatocytes at negative spontaneous curvatures and differential tensions $\alpha-\beta$, respectively, whereas the prolate and pear-like vesicles at positive curvatures resemble budded organoids found at positive differential tensions~({Supplementary Fig.~6}).}

The shapes shown in Fig.~\ref{F1}d as examples of active model organoids are characteristic of a given set of model parameters but they are not unique due to the stochasticity of active cell rearrangements. The resulting shape variability is illustrated by {Supplementary Fig.~5c}, which shows three branched organoids at $\alpha=0.7$ and $\beta=0.5$. All of these shapes have five unevenly articulated branches of somewhat different lengths and diameters. To systematically explore this variability, we simulate 300 instances at the four pairs of $\alpha$ and $\beta$ from Fig.~\ref{F1}d and we compute the corresponding distributions of reduced volumes. We find that shape variability differs among the four shapes, the spherical and the budded one being the least and the most variable, respectively~({Supplementary Figs.~5d-g}). The variability of the budded shape is likely due its location in the phase diagram: It lies close to the boundary between the budded and the branched morphologies, and thus in some instances the shape of the buds may be considerably more pronounced than in others. Nevertheless, throughout the phase diagram the standard deviation of the distribution never exceeds $\approx 0.04$, which shows that the reduced volumes of the organoid shapes are well-defined~({Supplementary Fig.~5b}). Note that since the budded and the stomatocyte morphologies appear at similar values of $\alpha+\beta$, their reduced volumes $v$ are similar as well~({Supplementary Fig.~5a}).

\subsection*{\label{sec:active_rearrangements}Active cell rearrangements}
\begin{figure}[htb!]
\centerline{\includegraphics{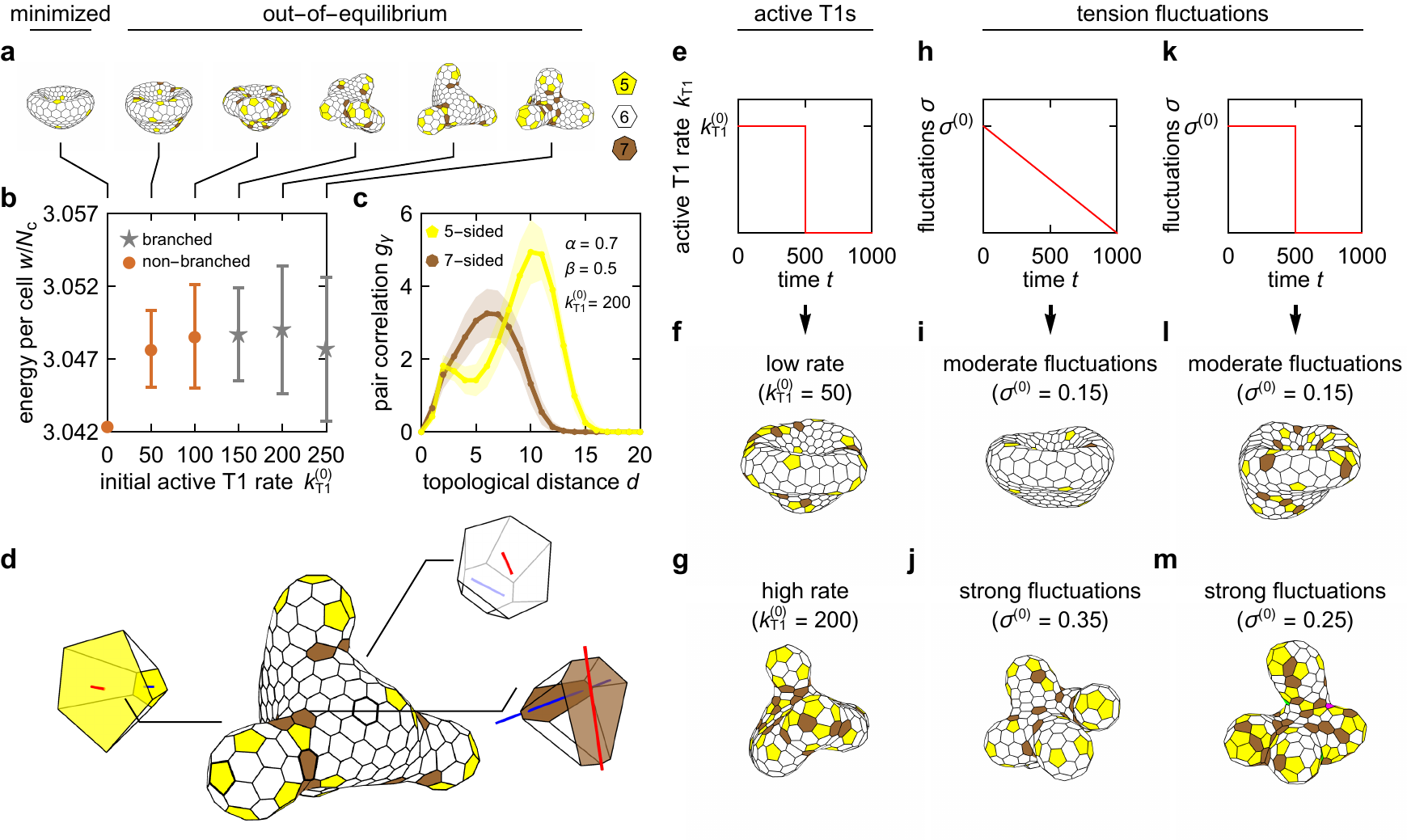}}
\caption{\textbf{High degree of active T1 transitions is necessary for formation of branched morphologies.} {\bf a}~Sequence of $\alpha=0.7,$ $\beta=0.5,$ $N_c = 300,$ and $v_{\rm lumen}={100}$  shapes with $k_{\rm T1}^{(0)}=0,50,100,150,200,$ and $250$. Cells are color-coded according to polygon class; pentagonal cells are yellow, hexagonal are white, and heptagonal are brown. {\bf b}~Energy of final organoid shapes as a function of the initial active T1 rate $k_{\rm T1}^{(0)}$. Error bars represent the standard deviation of energies over $100$ instances; at $k_{\rm T1}^{(0)}=0$ the mechanics is completely deterministic and the width of the error bar vanishes. {\bf c}~Distributions of topological distances between pairs of pentagons (yellow) and heptagons (brown) in the organoid at $k_{\rm T1}^{(0)}=200$ (fifth shape from the left in panel a. The symbols show the average over $300$ simulated instances, whereas the shaded areas represent the standard deviation. Solid lines guide the eye. In panel {\bf d}, the $k_{\rm T1}^{(0)}=200$ organoid is magnified and the full 3D shapes of representative 5-, 6-, and 7-coordinated cells are replotted with the long axes of their apical and basal sides indicated by blue and red lines, respectively; line lengths are proportional to the anisometry of the side in question (Methods). {Panels {\bf f} and {\bf g} show the model shapes at $k_{\rm T1}^{(0)}=50$ and 200 in case of a step-like temporal profile of the active T1 rate ({\bf e}). In panels {\bf i} and {\bf j}, we present the organoids obtained within a model where T1 transitions are generated by fluctuations in line tension with a linear temporal profile ({\bf h}), the former for $\sigma^{(0)}=0.15$ and the latter for $\sigma^{(0)}=0.35$; panels {\bf l} and {\bf m} show shapes obtained with a step-like variation of the magnitude of fluctuations ({\bf k}) for $\sigma^{(0)}=0.15$ and $0.25$, respectively.}}
\label{F2}
\end{figure}

Our active model organoids explore the energy landscape at a fixed cell number by actively rearranging the cells and dissipating the energy through friction with the environment~\cite{Krajnc18}. Due to the complexity of the energy landscape, there is no guarantee that the system reaches the global energy minimum by the end of simulation. Instead, due to the decreasing active T1 rate $k_{\rm T1}$, which gradually drops to zero in each simulation, it is plausible that the final small-organoid morphologies are trapped in local energy minima. If true, this would explain why the branched morphologies are only found in our simulations that include active cell rearrangements~(Fig.~\ref{F1}f). To test this, we first vary the initial active T1 rate $k_{\rm T1}^{(0)}$ at fixed $\alpha=0.7$ and $\beta=0.5$, which give a branched morphology at $k_{\rm T1}^{(0)} = 200$. Importantly, we find that the branched morphology develops only when $k_{\rm T1}^{(0)}$ is high enough and we find that it is associated with a significantly higher energy compared to the corresponding energy-minimized shape~(Figs.~\ref{F2}a and b). 
This result shows that the branched small organoids are inherently out-of-equilibrium shapes that require a sufficiently high degree of junctional activity to form. Furthermore, being trapped in the respective local energy minima, the shapes of branched organoids may significantly depend on the details of the relaxation protocol.

{
To challenge this possibility, we next replace the linear temporal profile of the active T1 rate by a step-like profile where the initial period with a constant T1 rate $k_{\rm T1}=k_{\rm T1}^{(0)}>0$ of duration $t = 500$ is followed by an equally long period with no T1 transitions~($k_{\rm T1}=0$). We again find that the branched morphologies only occur if $k_{\rm T1}^{(0)}$ is high enough during the initial period and the final shapes stay qualitatively the same as before~(Figs.~\ref{F2}e-g). We note that the branched morphology is formed soon after the beginning of the active phase as illustrated by the sequence of shapes in {Supplementary Fig.~7}. Thus neither the duration nor the precise temporal profile of the activity seem essential for the development of the branches, provided, of course, that the duration is long enough.

We further test how the formation of the branched morphology depends on the implementation of junctional activity by considering an entirely different numerical scheme of active junctional noise. Here T1 transitions arise directly from fluctuations in tensions along cell-cell junctions instead of the active-T1 scheme employed above. To this end, we extend the energy function [Eq.~(\ref{eq:energy1})] by a line-tension term
\begin{equation}
    w_\gamma=\sum_{i=1}^{\mathcal{E}}\gamma_i(t)\left( l_{ai} + l_{bi}\right)\>,
\end{equation}
describing fluctuations of tensions at cell-cell junctions, i.e., along the direction perpendicular to the apico-basal axis. Here the sum runs over all cell-cell junctions $\mathcal{E}$; $l_{ai}$ and $l_{bi}$ are lengths of the apical edge and the basal edge, respectively, that correspond to the $i$-th junction, whereas $\gamma_i(t)$ is the effective time-dependent line tension associated with these edges. We then assume that the fluctuations of the line tension $\gamma_i(t)$ are described by the Ornstein--Uhlenbeck process \cite{Curran17,Krajnc20} so that
\begin{equation}
    \frac{{\rm d}\gamma_i(t)}{{\rm d}t}= -\frac{1}{\tau}\gamma_i(t) +\xi_i(t)\>.
\end{equation}
Here $\tau$ is the time scale associated with the turnover of molecular motors (set to unity, $\tau=1$) and $\xi_i(t)$ is the Gaussian white noise with properties $\braket{\xi_i(t)}=0$ and $\braket{\xi_i(t)\xi_j(t')}=(2\sigma^2/\tau)\delta_{i,j}\delta(t-t')$, where $\sigma^2$ is the long-time variance of the tension fluctuation. If the magnitude of fluctuations $\sigma$ is sufficiently large, individual edges occasionally shrink to zero length, initiating a T1 transition (Methods). We analyze two variants of the tension-fluctuation scheme of T1 transitions. In the first one, $\sigma$ decreases linearly from some initial value $\sigma^{(0)}$ to zero (Figs.~\ref{F2}h-j) whereas in the second it has a step-like temporal profile where a period of constant tension fluctuations with $\sigma=\sigma^{(0)}$ is followed by an equally long period with $\sigma=0$ (Figs.~\ref{F2}k-m). In agreement with our initial scheme of active noise, branched morphologies are formed in both variants provided that the activity parameter $\sigma^{(0)}$ is large enough~(Figs.~\ref{F2}j and m).}

In order to better understand the formation of branches, we examine the in-plane tissue structure represented by the topology of the polygonal network of cell-cell junctions. We find that the branched shapes, obtained in the more active organoids, develop very different cell arrangements compared to the less active non-branched shapes~(Fig.~\ref{F2}a). In particular, the branched shapes contain many pentagonal and heptagonal cells, which accumulate at the tips of the branches and at their bases, respectively. The distribution of these cells within the organoid can be quantified by plotting their pair correlations as functions of the topological distance $d$ defined as the integer shortest path between two cells; $d=1$ for the nearest neighbors, $d=2$ for the next-nearest neighbors,  etc.~(Fig.~\ref{F2}c and Methods). For pentagons, we observe a bimodal distribution with a small peak at $d=2$, which corresponds to the distance between pentagons within the same branch, and a large peak at $d=10$, which corresponds to the distance between pentagons in different branches. In contrast, the distribution of heptagons, which are mostly located at the bases of the branches, consists of a single peak at $d\approx6$. Furthermore, since the distance between the bases is shorter than that between the tips, heptagons are more likely to be found closer to other heptagons than pentagons are to other pentagons. Note that these pair correlations are qualitatively similar in the budded shapes, which favor pentagons at the buds and heptagons in the saddle-like parts. In contrast, stomatocytes and spheres lack such features and are thus associated with unimodal distributions~({Supplementary Fig.~8}).

These results suggest that defects in cell packing, established through active cell rearrangements, may be crucial for the formation of the overall shape. We hypothesize that the observed characteristic distributions of pentagonal and heptagonal cells in the branched morphologies are established hand in hand with the shape itself. In particular, the high-activity conditions enable the formation of a temporarily biased distribution of defects in cell packing, in which pentagons and heptagons act as a seeds for positive and negative Gaussian curvature, respectively (Fig.~\ref{F2}d and {Supplementary Movie 3}). In turn, once the distribution of the Gaussian curvature is established across the organoid, pentagons become locked at the positive-Gaussian-curvature tips, whereas heptagons are bound to the saddles. As a result, this interplay between the in-plane cell organization and the Gaussian curvature preserves the overall organoid shape and thus causes it to become trapped away from the global energy minimum. In contrast, organoids that are never exposed to a high activity cannot develop the required distribution of pentagons and heptagons, which prevents the formation of multiple localized regions of positive Gaussian curvature necessary to develop branches. 

Surprisingly, even though the relation between the local Gaussian curvature of the surface and locations of the defects in packing of constituents is well known in general~\cite{Nelson87}, its potential relevance for organoid (and tissue) development has not yet been pointed out. Furthermore, it is possible that this mechanism plays an important role in cell turnover. Indeed, in one of the fastest renewing organs in mammals---the intestine---cells divide at the bases of the villi and are extruded at the tips~\cite{Buske11}. It could be that cell extrusion is initiated by the presence of cells with pentagonal bases. These cells may accumulate at the tips of the villi where their basal side is larger than their apical side simply because pentagonal cells are energetically favorable at positive Gaussian curvatures~(Fig.~\ref{F2}d and {Supplementary Fig.~9a}). With their truncated-pyramid shape, such cells could conceivably more easily delaminate from the tissue due to mechanical reasons. In contrast, cells located at the bases of the villi where the Gaussian curvature is negative are preferentially heptagonal. In these cells, the local saddle-like shape of the tissue is consistent with somewhat anisometric shapes of the apical and the basal side, with the two long axes roughly perpendicular to each other~(Fig.~\ref{F2}d and {Supplementary Figs.~9b-d}). As the areas of the two sides should be roughly the same, cells of such a form are stabilized against extrusion and act as anchors. Overall, this hypothesis suggests that the curvature-dependent extrusion rate, observed in intestinal epithelia~\cite{Hannezo11}, could be to a large extent of mechanical origin. Furthermore, our findings suggest that the basis of this mechanism in organoids may be established spontaneously during shape formation.

\subsection*{\label{sec:continuum}Curvature-thickness coupling}
To better understand the obtained shapes, we recast the cell-level energy Eq.~(\ref{eq:energy1}) in the continuum limit following Ref.~\cite{Krajnc15}. We approximate cell shape by a truncated right square pyramid, a body with in-plane isometry, imposing a fluid-like response to local shear stress. Cell shape is parametrized by the local principal curvatures $c_1$ and $c_2$ of the midplane as well as by tissue thickness $h$, i.e., cell height, measured in units of $1/V_c^{1/3}$ and $V_c^{1/3}$, respectively ({Supplementary Note 1}). In the first-order approximation, the dimensionless elastic energy per unit area reads
\begin{equation}
\frac{{\rm d} w}{{\rm d} a}=(\alpha+\beta)+2\left[1-\frac{(\alpha-\beta)^2}{4}\right]\sqrt{h^3}+\frac{\sqrt{h}}{8}\left[c_1+c_2-2\sqrt{h}(\alpha-\beta)\right]^2+\left(\frac{\alpha+\beta}{4}+\frac{\sqrt{h^3}}{12}-\frac{1}{4\sqrt{h^3}}\right)h^2c_1c_2.
\label{eq:bendingenergy}	
\end{equation}
The first two terms together represent the surface tension whereas the third and the fourth term are the local and the Gaussian bending energy per unit area, respectively. Note that the thickness $h$ in Eq.~(\ref{eq:bendingenergy}) is an independent variable: While it is constant in a flat tissue at $\alpha=\beta$, in the nontrivial shapes $h$ is coupled to the local curvature. As a result, the surface tension varies along the surface and so do the local bending modulus $\sqrt{h}/8$, the spontaneous curvature $c_0=2\sqrt{h}(\alpha-\beta)$, and the Gaussian modulus [i.e., the last term in Eq.~(\ref{eq:bendingenergy}) divided by the Gaussian curvature $c_1c_2$]. These features render the tissue energy functional quite different from the usual surface and bending energies of solid or fluid membranes.
\begin{figure}[ht]
	\centerline{\includegraphics{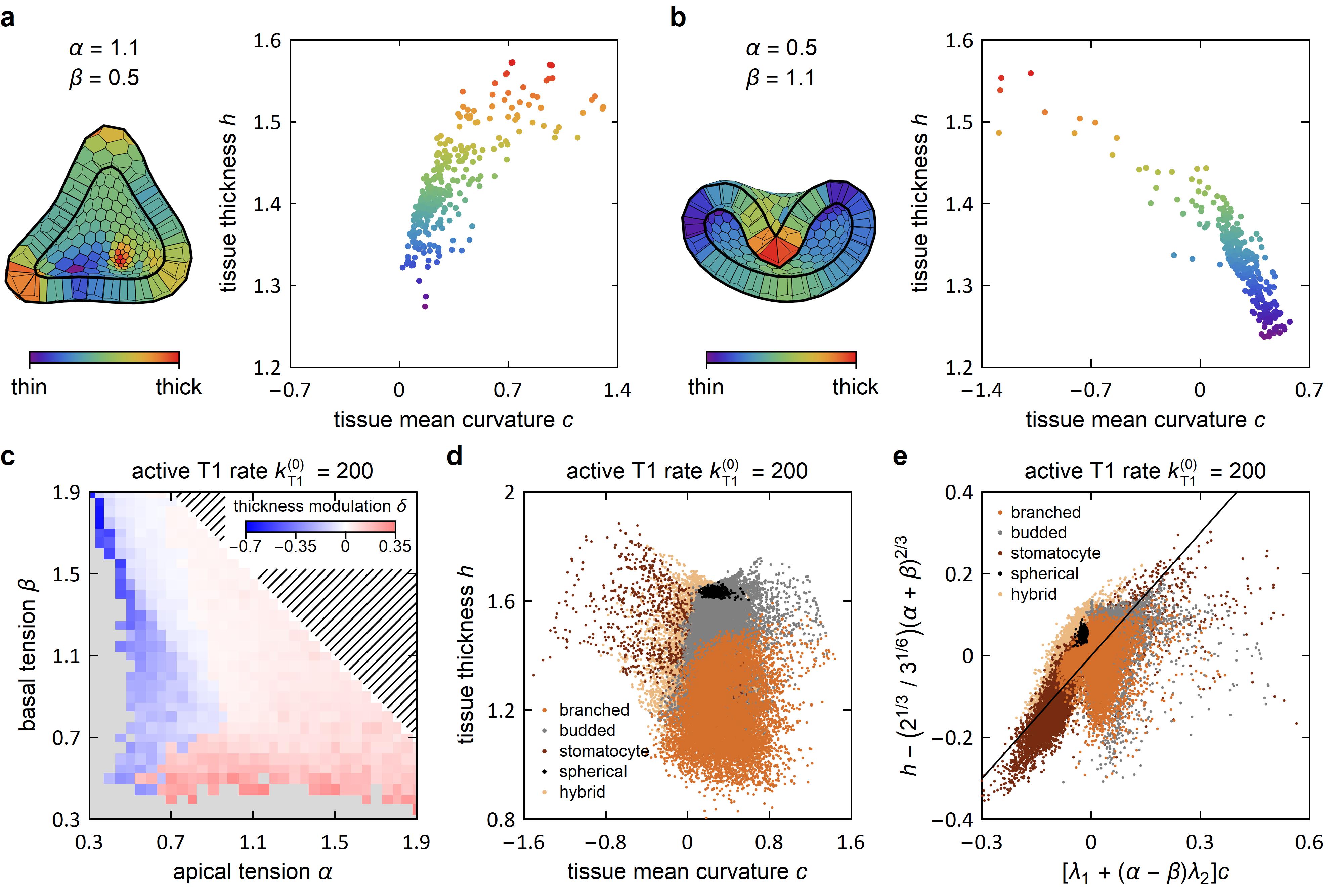}}
	\caption{\textbf{Thickness-curvature coupling arises due to apicobasal polarity.} {\bf a}, {\bf b}~Thickness-curvature portraits of the active T1 ($k_{\rm T1}^{(0)}=200$) budded shape with $\alpha = 1.1,\>\beta = 0.5$~(panel a) and of a stomatocyte shape with $\alpha = 0.5,\>\beta = 1.1$~(panel b); each cell is represented by a point. In the cutaway view of the model organoids, tissue thickness is represented using color code which is also used for the points in the two diagrams. {\bf c}~Modulation of tissue thickness $\delta$ in shapes in the $(\alpha,\beta)$-plane for $k_{\rm T1}^{(0)}=200$. The gray region represents the non-physical regime whereas the hatched region denotes the spherical shapes at $\alpha+\beta>2.6$. {
	{\bf d}~The 55500 cell-by-cell $h(c)$ datapoints from the 185 model shapes from panel c with $\alpha+\beta<1.9$. {\bf e}~Datapoints from panel d collapsed by rescaling described in the text; also plotted is the line of identity.}}
	\label{F3}
\end{figure}

The coupling between the mean curvature and the thickness $(\beta-\alpha)(c_1+c_2)h/2$, contained in the local bending term, is the most important of the above effects. In the more elaborate shapes, this coupling is quite pronounced as illustrated by Figs.~\ref{F3}a and b showing the thickness-curvature portraits of a budded and a stomatocyte shape, respectively. { In these two diagrams, each cell is represented by a point indicating the mean curvature $c=(c_1+c_2)/2$ and thickness $h$ of the tissue at this cell (Methods).} The positive correlation between thickness and curvature is evident in the budded shape (Fig.~\ref{F3}a) as is the negative correlation in the stomatocyte (Fig.~\ref{F3}b). In the budded $\alpha=1.1$, $\beta=0.5$ shape with a positive spontaneous curvature $c_0$ (i.e.~a spontaneous curvature with the same sign as the curvature of a sphere), the taller cells are located at the buds where the mean curvature is largest~(Fig.~\ref{F3}a), whereas in the $\alpha=0.5$, $\beta=1.1$ stomatocyte where $c_0<0$, they are concentrated at the invagination~(Fig.~\ref{F3}b).

To quantify the curvature-thickness coupling throughout the phase diagram, we introduce the relative tissue thickness modulation $\delta = p \cdot \Delta h / \overline{h}$, where $\Delta h$ is the amplitude of thickness modulation whereas $\overline{h}$ is the average cell height in the model organoid; the prefactor $p$ takes the value of $+1$ if curvature and thickness are correlated~(Fig.~\ref{F3}a) or $-1$ if they are anticorrelated~(Fig.~\ref{F3}b). We find that the thickness modulation is sizable in all non-spherical shapes, with $|\delta|$ reaching about 0.65 in stomatocytes and 0.35 in budded and branched shapes~(Fig.~\ref{F3}c). Consistent with the continuum theory~[Eq.~(\ref{eq:bendingenergy})], the sign of $\delta$ agrees with the sign of the modulus of the curvature-thickness coupling $\alpha-\beta$.

{ 
We can further analyze the coupling between curvature and thickness by fitting the $h(c)$ plots for individual model organoids by linear functions. We then plot the vertical intercept $n$ of these functions against tissue tension $\alpha+\beta$ and the slopes $k$ against the differential tension $\alpha-\beta$ for all 185 model organoid shapes discussed here~({Supplementary Fig.~10}). We find that the vertical intercept $n$ can be described by the above analytical solution for hexagonal cells in a flat epithelium $n\approx\left (2^{1/3}/3^{1/6}\right )(\alpha+\beta)^{2/3}$, whereas the slope $k$ can be approximated by $\lambda_1+(\alpha-\beta)\lambda_2$ where $\lambda_1=-0.060\pm0.004$ and $\lambda_2=0.45\pm0.01$ are fitting parameters. (These two observations only apply to organoids with $\alpha+\beta<1.9$. Beyond this threshold, organoids approach a spherical shape and their midplane area is defined by the enclosed volume rather than by $\alpha$ and $\beta$.) We can then rescale $h\rightarrow h - \left (2^{1/3}/3^{1/6}\right )(\alpha+\beta)^{2/3}$ and $c\rightarrow \left[\lambda_1+(\alpha-\beta)\lambda_2\right]c$, which allows us to collapse all of the $h(c)$ scatter plots for the model organoids at $k_{\rm T1}^{(0)}=200$ with $\alpha+\beta<1.9$ (Figs.~\ref{F3}d-e); the same can also be done for the organoids at $k_{\rm T1}^{(0)}=0$ in Fig.~\ref{F1}e.

While the modulation of cell height is often attributed to cell differentiation~\cite{Eiraku11,Kuwahara15}, our model shows that it can appear even in tissues of identical cells where it is a telltale sign of mechanical apico-basal polarity.} Similar shape features are often observed in the folded morphologies in a variety of tissues in both vertebrates and invertebrates~\cite{Storgel16}.

\section*{\label{sec:Discussion}Discussion}
\begin{figure}[htb!]
\centerline{\includegraphics{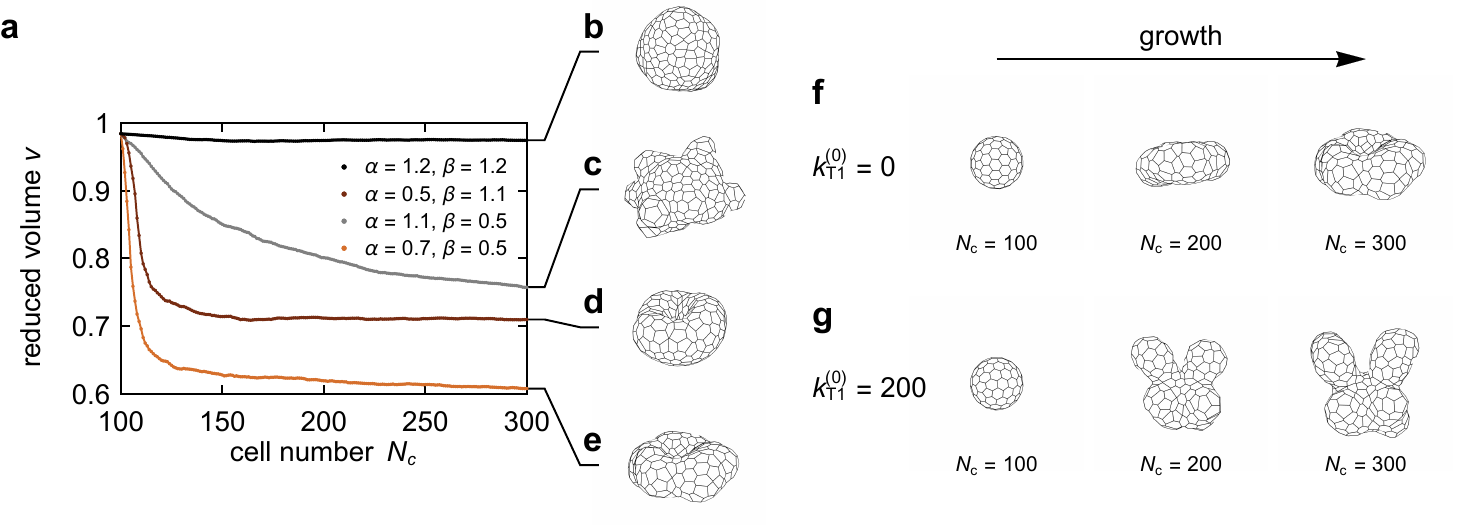}}
\caption{\textbf{Model of growing organoids.} {\bf a}~Reduced volume of growing organoids as a function of cell number. {\bf b}-{\bf e}~Organoid shapes at $N_c=300$ and $(\alpha,\beta)=(1.2,1.2),\>(1.1,0.5),\>(0.5,1.1),$ and $(0.7,0.5)$, respectively, based on 25 instances. {{\bf f}~Growing organoid with $(\alpha,\beta)=(0.7,0.5)$ at 100-, 200-, and 300-cell stage in absence of active T1 transitions. {\bf g}~Growing organoid at the same $\alpha$ and $\beta$ but  with active T1 transitions.} }
\label{FDiv}
\end{figure}

While our model yields a diverse catalog of possible small-organoid morphologies, their formation in real systems may rely on processes other than surface tension and active cell rearrangements. In particular, the shape usually develops while cells in an organoid grow and divide. To test how shapes are established during growth and how they compare to those predicted by our original approach, we generalize our model by including volumetric cell growth and cell division~(Methods). 
We simulate growing organoids at fixed cell-growth and cell-division rates, $\tau_d=2000$ being the expected time until a cell next divides, whereas the lumen volume follows Eq.~(\ref{eq:rescale}) (Methods). We start from a sphere of 100~cells and we monitor the growing organoid until the number of cells reaches 300. We first compute the shapes in absence of activity (i.e, $k_{\rm T1}^{(0)}=0$, where only spontaneous T1 transitions are allowed) and find that in terms of their final reduced volumes, the thus obtained shapes agree well with those from our original model~(Fig.~\ref{FDiv}a and {Supplementary Figs.~5d-g}). The spherical, budded, and stomatocyte shapes all develop their characteristic features, although they are somewhat less regular than in their non-growing counterparts due to a different relaxation mechanism~(Figs.~\ref{FDiv}b-d and {Supplementary Movie~4}). In contrast, the $\alpha=0.7$, $\beta=0.5$ shape does not develop branches although its reduced volume is similar to that of the branched shapes obtained in active organoids at a fixed cell number (Figs.~\ref{FDiv}a, e, and f). These results show that cell-scale activity due to cell divisions is not sufficient to form branched morphologies.

{To check if branching can be recovered in growing organoids by including junctional activity, we combined cell division with the previously used active T1 transitions with a linearly decreasing T1 rate [$k_{\rm T1}(t) = k_{\rm T1}^{(0)} (t_{\rm max} - t) / t_{\rm max}$, where $k_{\rm T1}^{(0)}=200$ and $t_{\rm max}=1000$; this set of parameters causes the active T1 rate to reach zero at a time when the organoid contains about 160 cells]. As shown by the shapes in Fig.~\ref{FDiv}g and {Supplementary Movie~5}, this level and duration of active T1 transitions are sufficient for the branched morphology to develop, which is in agreement with our observation that organoids require a certain degree of junctional activity to branch~(Figs.~\ref{F2}a~and~b).}

Our results show that small organoids can form a variety of shapes relying on a simple collective mechanics of non-differentiated cells, coupled through a fluid-like organoid interior. We explore a scenario where a complex shape forms due to a mismatch between the organoid's preferred surface area and the volume of the enclosed lumen. The four characteristic morphologies---spherical, stomatocyte, budded, and branched---are distinguished by the reduced volume. We also develop a continuum theory and use it to explain the fine features of the shapes (specifically, the organoid-scale cell-height modulation) which could be measured {\it in vitro} to extract the apico-basal surface-tension polarity directly from the shape. Overall, our model provides clues for the control of organoid shapes in experiments. This can be done by changing the relative apical and basal tensions of cells, which should depend on the biochemical composition of the environment (i.e., the contents of the lumen and the medium). These tensions are best viewed as effective concepts encompassing all membrane-related effects and processes that shape the cells, and may be altered in several ways including, e.g., the activation of the complex cascade leading to contraction of the apical network~\cite{Lecuit07}. Our results also show that the organoid shape may be affected by suppression or promotion of junctional activity and, importantly, its effect on the topology of the tissue represented by the polygonal tiling seen on the basal or apical plane. 
A systematic assessment of the different experimental options could conceivably expand the toolbox of the organoid technology and tissue engineering~\cite{Griffith02}. 

A natural extension of this work would be to include planar cell polarity, which can induce tubulogenesis and thus branching~\cite{Nissen18}, and cell differentiation, which could contribute to self-organization of cells within organoids through differential adhesion as well as to programmed shape formation through spatial modulation of the apico-basal polarity. To better describe the typical experimental setup in growing organoids, the model could consider tissue growth on curved solid substrates { and include viscous dissipation at the tissue-matrix interface, which may promote buckling and branching~\cite{Okuda15}}. In such a case, other cell-scale active processes, e.g., cell motility, could contribute to the collective cell dynamics and the formation of the overall organoid morphology. 

\section*{Methods}

\subsection*{\label{ap:implementation}Implementation}

Simulations of organoid shapes are performed within vertX3D, a package of C++ routines that implement the 3D vertex model of epithelial tissues. The initial configuration of cells is a spherical shell enclosing the lumen volume $v_{\rm lumen}$. These configurations are constructed by finding the dual of the convex hull to the minimal-energy solution of the Thomson problem on a sphere~\cite{Wales06}. The dual determines the positions of the apical vertices, whereas the positions of the basal vertices are calculated by displacing their apical counterparts radially by $[3(v_{\rm lumen} + N)/(4\pi)]^{1/3}-[3v_{\rm lumen}/(4\pi)]^{1/3}$. To correct the initial cell volumes, which do not agree exactly with the preferred volumes, the organoids are first relaxed at $k_{\rm T1} = 0$ for a short time before the active T1 rate is set to $k_{\rm T1}^{(0)}$. The preferred cell volumes and the lumen volume are enforced by auxiliary harmonic terms in the energy; the modulus of these terms is very large ($100$ unless stated otherwise).

{ In the active-T1 scheme, the T1 transitions are initiated as described in the main text; in the fluctuating tension scheme, they are only initiated if the length of a junction expressed in units of $V_{\rm cell}^{1/3}$ falls under 0.01.} In both cases, T1 transitions are applied by changing the connectivity of the apical network, and then updating the basal network and the lateral sides accordingly.  During the transition, the two apical vertices defining the edge that undergoes the T1 event are first moved to the average of their previous positions. The four-way-rosette arrangement is left for a short time interval of $2\times10^{-3}$ before it is resolved. The resolution happens by separating the new pair of apical vertices by a distance 0.0005 in a symmetric fashion. 

{ In organoids with no active T1 transitions ($k_{\rm T1}^{(0)}=0$) simulations are run until $t=2000$, i.e., twice as long as in organoids with junctional activity; this is because the stomatocyte shapes at $\alpha\approx\beta$ require a longer time to fully develop.}
	
In our vertex model, steric repulsion between cells is not implemented and self-overlapping shapes are therefore possible. During post-processing, we check each obtained final shape and discard all that self-overlap; temporary overlaps may occur during the simulation.

When analyzing the obtained organoid shapes, we compute cell height (i.e., tissue thickness) reported in Fig.~\ref{F3} as the distance between the centroids of the apical and the basal side, and we approximate the mean curvature of each cell by that of a truncated cone with the same apical area, basal area, and height.

When considering growing organoids, cells that are otherwise subject to the fixed-volume constraint enter the growth perido with a constant probability $\textup{d}P/\textup{d}t=1/\tau_d$, where $\tau_d=2000$ is the characteristic time scale of division events. During the growth period, the cell volume is increased linearly in time according to $\textup dV/\textup dt=1/\tau_g$, where $\tau_g=1$ is the cell-growth time scale. As soon as the volume is doubled, the cell is divided with the mitotic plane connecting a random pair of opposing lateral sides. Starting with a spherical organoid of 100 cells, we let the cells divide while increasing the preferred lumen volume according to Eq.~(\ref{eq:rescale}), substituting $N_c$ by the total volume of the tissue. 

{
\subsection*{\label{ap:analytical}Volume and active T1 rate rescaling}
When comparing morphologies with different cell numbers $N_c$ in {Supplementary Fig.~3}, the active T1 rate $k_{\rm T1}$ and lumen volume $v_{\rm lumen}$ must be appropriately rescaled for the comparison to be relevant. As the probability for an individual edge to undergo a T1 transition is $k_{\rm T1} \delta t / \mathcal{E}$, the most relevant comparison is between model organoids where an individual cell-cell junction has the same probability of undergoing a T1 at both $N_c$ in question. We therefore rescale $k_{T1}\rightarrow k_{T1} \mathcal{E}_{N_c} / \mathcal{E}_{300}$; here $\mathcal{E}_{N_c}$ is the number of cell-cell junctions in an organoid of $N_c$ cells and $\mathcal{E}_{300}$ is their number in a 300-cell organoid. The same rescaling is also used when considering growing organoids (Fig.~\ref{FDiv}) so that edges have an equal probability of undergoing an active T1 transition regardless of the number of cells.

When rescaling lumen volumes, we choose them such that the model organoids have the same $\alpha$, $\beta$ threshold for the spherical shape, which can be estimated as follows. As before, we approximate cell height by $h_0=\left (2^{1/3}/3^{1/6}\right )(\alpha+\beta)^{2/3}$. The threshold for a spherical shape is then the value of tissue tension $\alpha+\beta$ at which a spherical shell of thickness $h_0$ around the lumen of volume $v_{\rm lumen} (N_c)$ has volume $N_c$. We then equate the $\alpha+\beta$ values at the given target $N_c$ and the reference $N_c = 300$, giving the condition
\begin{equation}\label{eq:rescale}
    \left[v_{\rm lumen} (N_c)+N_c\right]^{1/3}-\left[v_{\rm lumen} (N_c)\right]^{1/3}=\left[v_{\rm lumen} (N_c)+300\right]^{1/3}-\left[v_{\rm lumen}(300)\right]^{1/3}.
\end{equation}
Here $v_{\rm lumen}(300)=100$ is the lumen volume at 300 cells, whereas $v_{\rm lumen}(N_c)$ is the lumen volume at the target cell number $N_c$. This equation can trivially be solved numerically for $v_{\rm lumen}(N_c)$.
}
\subsection*{\label{ap:pair_correlation}Topological pair correlations}
The topological pair correlation function $g_\gamma(d)$ encodes how many $\gamma$-sided cells may on average be expected at a topological distance $d$ from another $\gamma$-sided cell. It is defined as
\begin{equation}
g_\gamma(d) = \left< \frac{1}{n_\gamma} \sum_{i = 1}^{n_\gamma} \mathcal{N}_{\gamma i}(d)\right>.
\label{eq:pairwisecorrelation}	
\end{equation}
Here the sum runs over all $\gamma$-sided cells and $\mathcal{N}_{\gamma i}(d)$ is the number of $\gamma$-sided cells at a topological distance $d$ from the $i$-th $\gamma$-sided cell. The normalizing prefactor $n_\gamma$ is the total number of $\gamma$-sided cells in the organoid. The expression in the brackets is averaged over 300 simulation runs.

\subsection*{\label{ap:anisotropy}Shape anisometry}
The long axes of the apical and basal sides of cells in Fig.~{\ref{F2}d} and {Supplementary Fig.~9} are obtained by diagonalizing the gyration tensor computed from the vertices of a side. The anisometry factor of a cell side $\kappa$ is given by $\kappa=(g_1-g_2)/(g_1+g_2)$, where $g_1$ and $g_2$ are the largest and the second largest eigenvalue of the gyration tensor, respectively; as the cell sides are slightly non-planar, the third eigenvalue is finite rather than 0 but still much smaller than $g_1$ and $g_2$.

\section*{Data availability}

The data containing the computed organoid shapes is available from the corresponding author on reasonable request.

\section*{Code availability}

The codes used to compute the organoid shapes is available from the corresponding author on reasonable request.


\section*{Additional Information}
\textbf{Supplementary Information} accompanies this paper.
\\
\noindent\textbf{Competing interests:} The authors declare no competing interests.

\section*{Author Contributions}

P.Z. conceived the project; J.R. and M.K. carried out the numerical work; J.R. analyzed the data; J.R., M.K., and P.Z. wrote the paper.

\section*{Acknowledgments}

We thank J.~Heuberger, S.~Hopyan, and S.~Svetina for helpful discussions. MK thanks Stas Shvartsman for support. The authors acknowledge the financial support from the Slovenian Research Agency (research core funding No.~P1-0055 and projects No.~J2-9223 and No.~Z1-1851). 

\end{document}


\maketitle
\thispagestyle{empty}
\setcounter{page}{0}
\newpage

\section*{Supplementary Figures}
	\begin{figure}[H]
		\centering
		\includegraphics{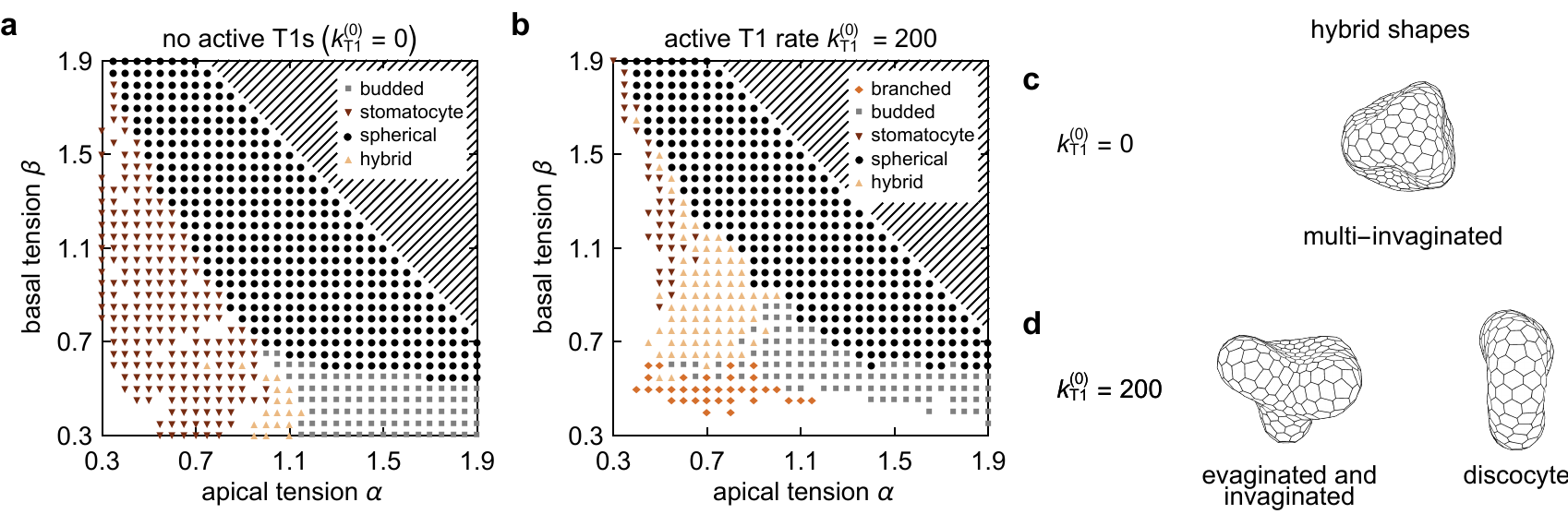}
		\caption{\textbf{Morphological phase diagrams of model organoids.} {\bf a}~Raw data at $k_{\rm T1}^{(0)} = 0$. Shapes with $\alpha + \beta >2.6$ (hatched region) are spherical and not shown for clarity. Self-intersecting and otherwise non-physical shapes at small $\alpha$ or $\beta$ are not included~(Methods). {\bf b}~Raw data at $k_{T1}^{(0)}=200$. The hatched region again denotes spherical shapes with $\alpha + \beta >2.6$, and self-intersecting and otherwise non-physical shapes are not shown. {\bf c}~Example of a hybrid shape with multiple invaginations at $N_c = 100$, $v_{\rm lumen} = 100$, and $k_{\rm T1}^{(0)} = 0$, with $\alpha = 1.1$ and $\beta = 0.4$. {\bf d}~Examples of hybrid shapes at the same $N_c$ and $v_{\rm lumen}$, but with $k_{\rm T1}^{(0)} = 200$. Here hybrid shapes appear at $\alpha \approx \beta$  and often feature either both evaginations and invaginations (left; $\alpha = 0.7$ and $\beta = 0.65$) or multiple invaginations, giving rise to a discocyte-like shape (right; at $\alpha = 0.55$ and $\beta = 0.7$). }
	\end{figure}
	\newpage
	
	\begin{figure}[H]
		\includegraphics{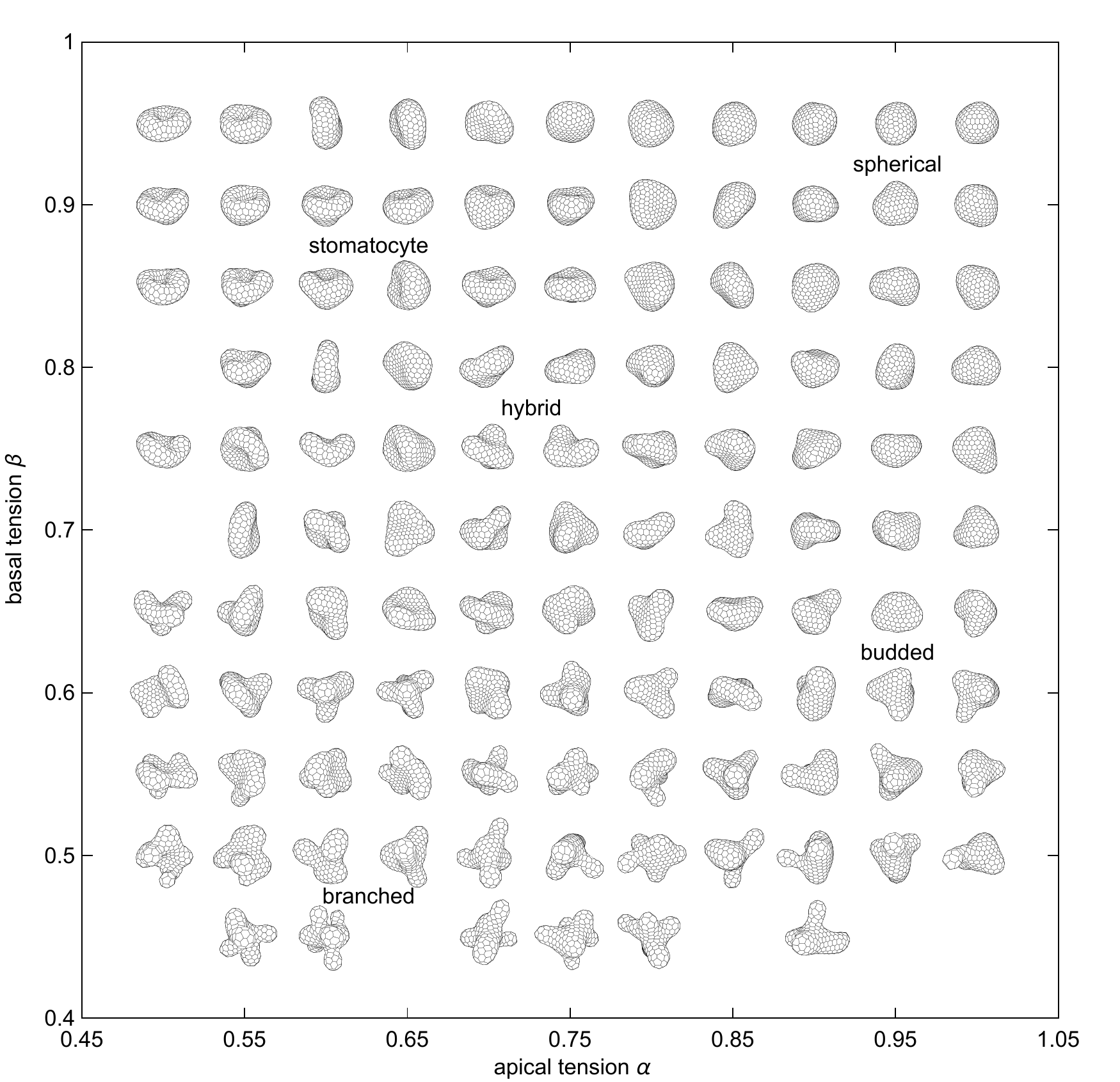}
		\caption{\textbf{Diagram of small-organoid shapes.} Small-organoid shapes from the bottom-left region of the phase diagram in Fig.~1f of the main text with $k_{\rm T1}^{(0)}=200$, showing various budded, branched, stomatocyte, and hybrid shapes as well as some spherical shapes.}
	\end{figure}
	\newpage
	
	\begin{figure}[H]
		\centering
		\includegraphics{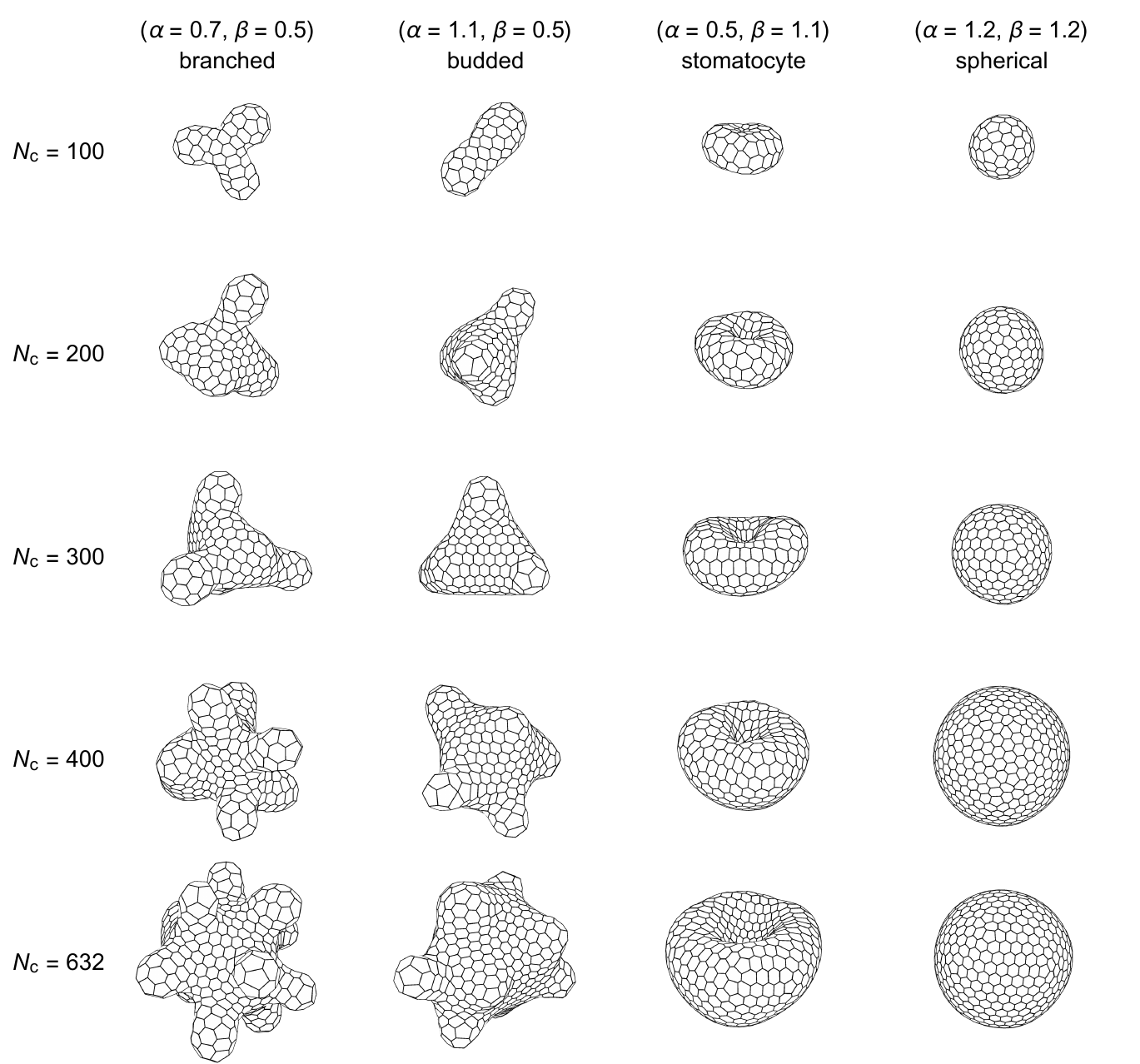}
		\caption{\textbf{Comparison of organoids containing different numbers of cells.} Organoid morphologies at $k_{\rm T1}^{(0)} = 200 \times \mathcal{E}_{N_c}/\mathcal{E}_{300} $ for $N_c=100, 200, 300, 400,$ and 632 at parameters $(\alpha=0.7,\>\beta=0.5)$, $(\alpha=1.1,\>\beta=0.5)$, $(\alpha=0.5,\>\beta=1.1)$, and $(\alpha=1.2,\>\beta=1.2)$; $\mathcal{E}_{N_c}$ is the number of cell-cell junctions of a shape at a given number of cells $N_c$ (Methods). Given an appropriate rescaling of lumen volumes, the same model parameters give the same type of morphology.}
	\end{figure}
	\newpage
	
	\begin{figure}[H]
		\centering
		\includegraphics{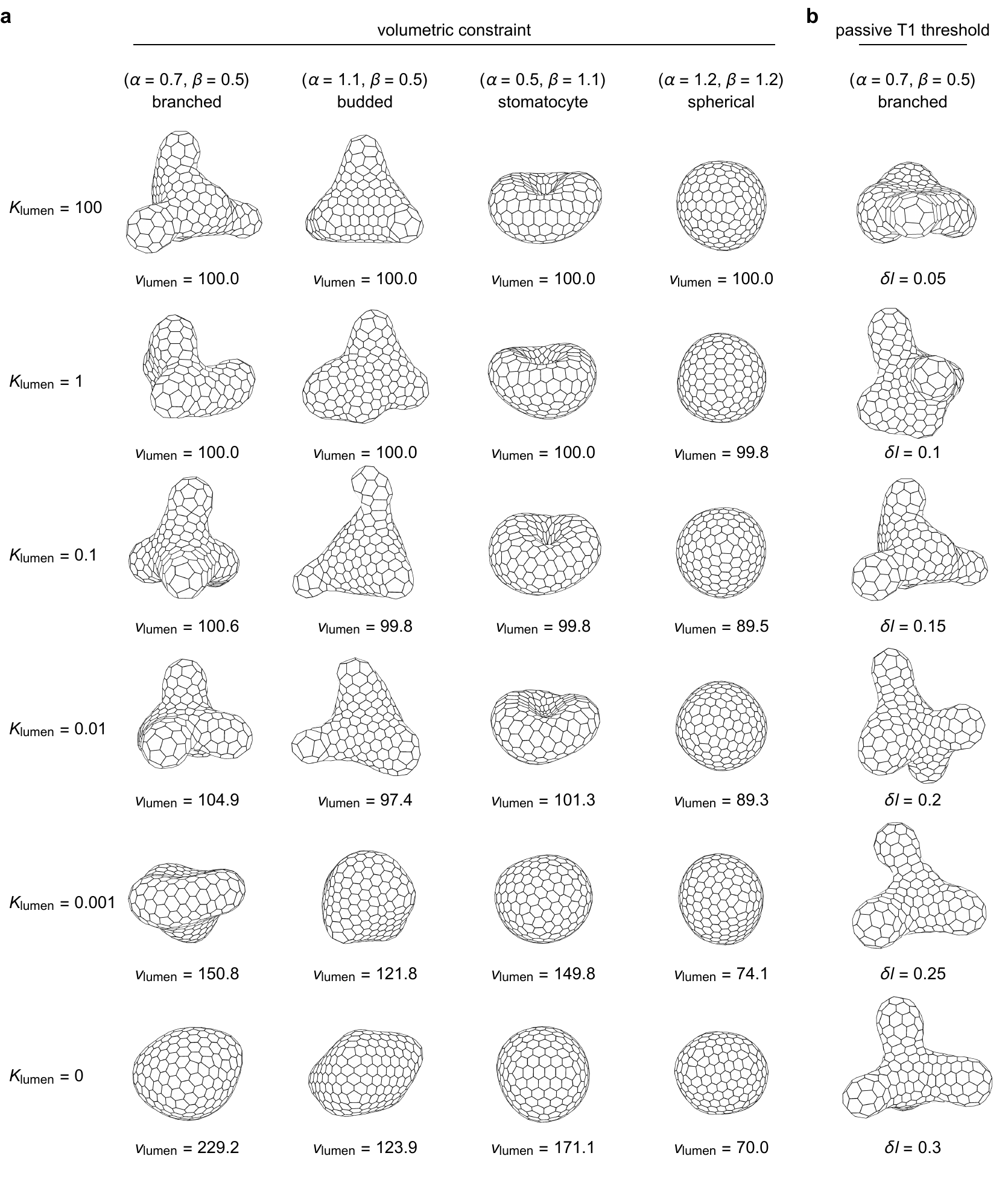}
		\caption{\textbf{Influence of modulus of lumen volumetric term and passive T1 transition threshold length of model organoid morphologies.} {\bf a}~Organoid morphologies at moduli of the auxiliary lumen volumetric term $K_{\rm lumen}$ equal to $100,1,0.1,0.01,0.001$ for four sets of $\alpha$ and $\beta$. The preferred lumen volume equals 100 in dimensionless units; the actual lumen volume is given under each organoid shape. If $K_{\rm lumen}$ is too small, the model organoids are devoid of clear morphological features. {\bf b}~Organoid morphologies at $\alpha=0.7,\ \beta=0.5$ at different values of the threshold for passive T1 transitions $\delta l$. Branches only develop at sufficiently large $\delta l$.}
	\end{figure}
	
	\begin{figure}[H]
		\centering
		\includegraphics{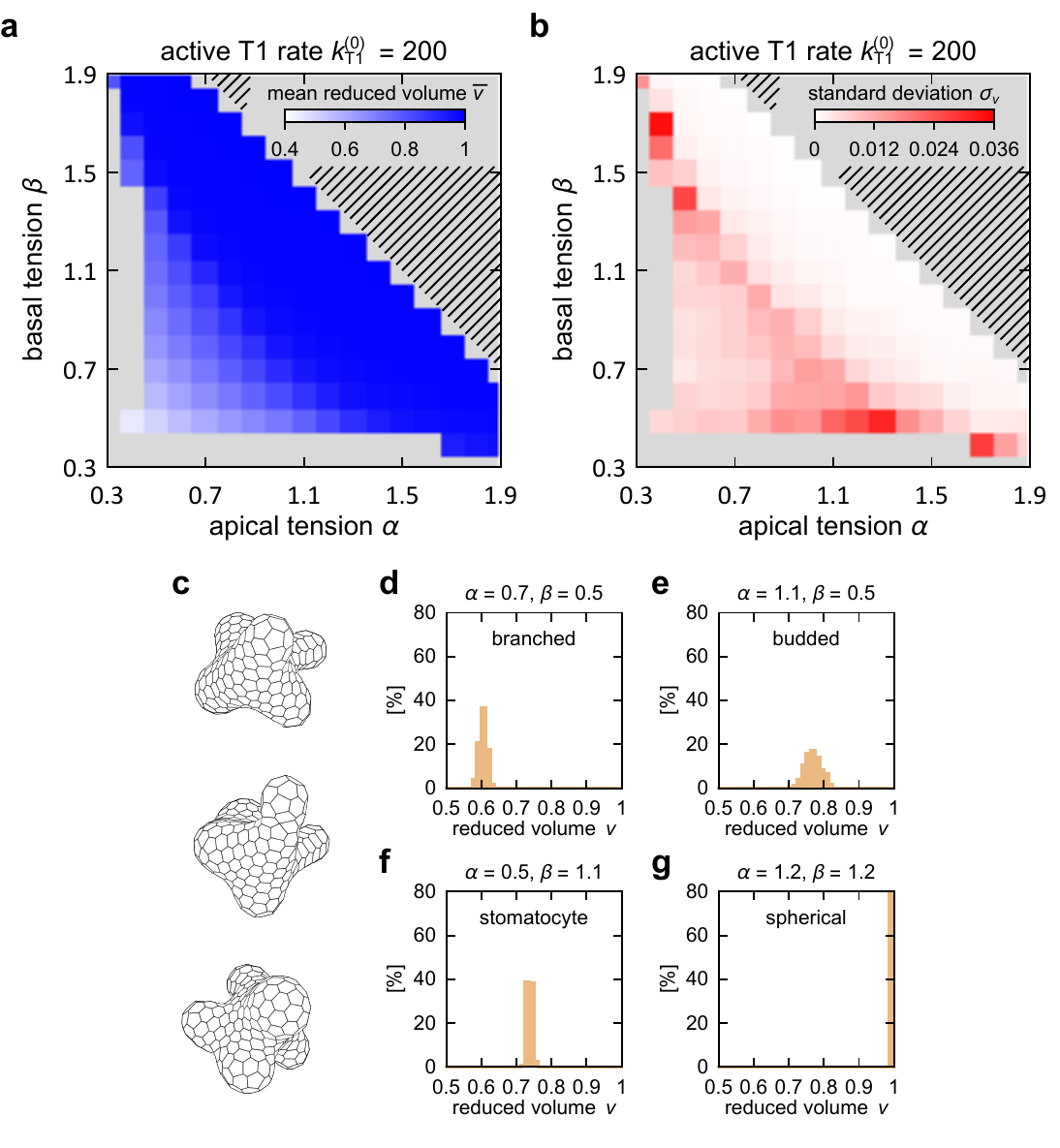}
		\caption{\textbf{Analysis of small-organoid reduced volumes in active model organoids.} 
		{\bf a}~Mean reduced volume from 35 simulation runs. The reduced volume of a shape depends almost exclusively on the tissue tension $\alpha+\beta$ and is virtually independent of the differential tension $\alpha-\beta$. {\bf b}~Standard deviation of the reduced volumes from the 35 simulation runs, showing that the reduced volume of the shape at a given $\alpha$ and $\beta$ is well-defined. {{\bf c}~Three instances of a branched organoid at $\alpha = 0.7,$ $\beta = 0.5,$ $N_c=300,$ $v_{\rm lumen}={100}$, and $k_{\rm T1}^{(0)}={200}$. {\bf d}-{\bf g} Distributions of reduced volumes for 300 instances of shapes at $\alpha$ and $\beta$ corresponding to the branched, budded, stomatocyte, and spherical morphologies in Fig.~1d of the main text}.
		}
	\end{figure}
	
	\begin{figure}[H]
		\centering
		\includegraphics{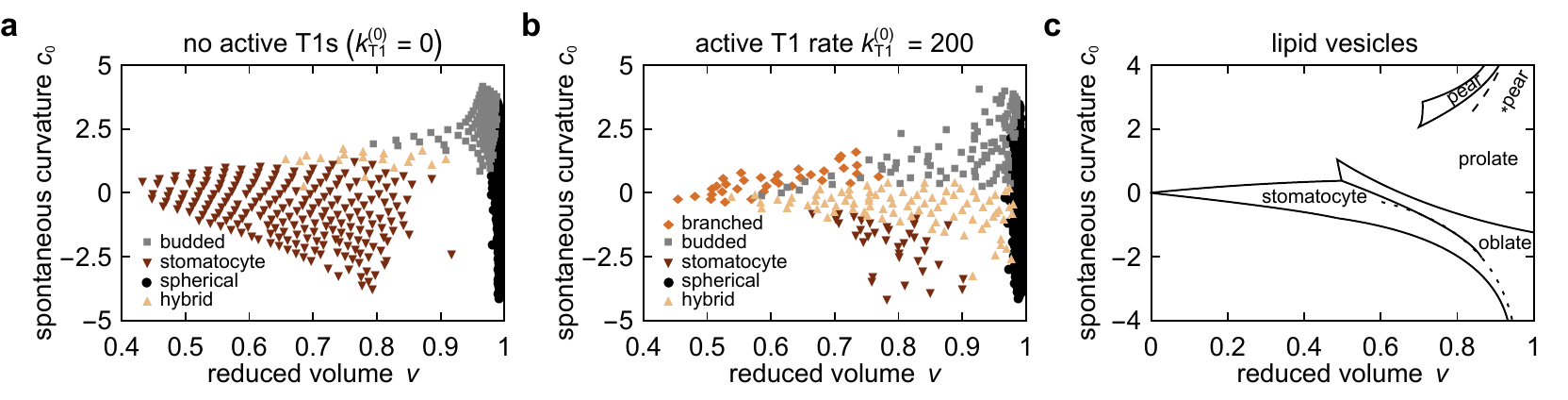}
		\caption{\textbf{Comparison between small-organoids and lipid vesicles.} {\bf a}~Phase diagram of organoids with no active T1 transitions ($k_{\rm T1}^{(0)}=0$) from Supplementary Fig.~1a plotted in the space spanned by the reduced volume $v$ [Eq.~(2) of the main text] and spontaneous curvature $c_0 = 2 (\alpha + \beta) ^{1/3}(\alpha-\beta)$ [Eq.~({\ref{c0}}) of Supplementary Note 1]. {\bf b}~Phase diagram of $k_{\rm T1}^{(0)}=200$ organoids from Supplementary Fig.~1b plotted in the $(v,c_0)$-plane. {\bf c} Phase diagram of spontaneous-curvature bilayer vesicles (adapted from Ref.~\cite{Seifert91}). 
		The trumpet-like arrangement of points in panel b follows from the fact that in any stable shape, both $\alpha$ and $\beta$ are restricted in magnitude: If one of them (say $\alpha$) is negative, its absolute value should not exceed a certain threshold or else the energy of individual cells is unbounded from below. By considering completely flattened squamous cells, we find that $|\alpha|<\min(1/2,\beta)$, which in turn imposes a restriction on the magnitude of $c_0$. In the numerically obtained closed organoid shapes, the actual restrictions are somewhat different from those obtained by analyzing the shape of individual cells but the qualitative form of the physically relevant domain is still trumpet-like although not uniformly populated. In panel a, the trumpet form of the domain is primarily due to the mapping of the data from Supplementary Fig.~1a; most  $k_{\rm T1}^{(0)}=0$ shapes are unaffected by the above restrictions, which are less limiting than in their generally more elaborated $k_{\rm T1}^{(0)}=200$ counterparts. The locations of the different morphologies in panels a and b are reasonably similar to those of their vesicle analogs in the phase diagram in panel~c.}
	\end{figure}
	\begin{figure}[H]
		\centering
		\includegraphics{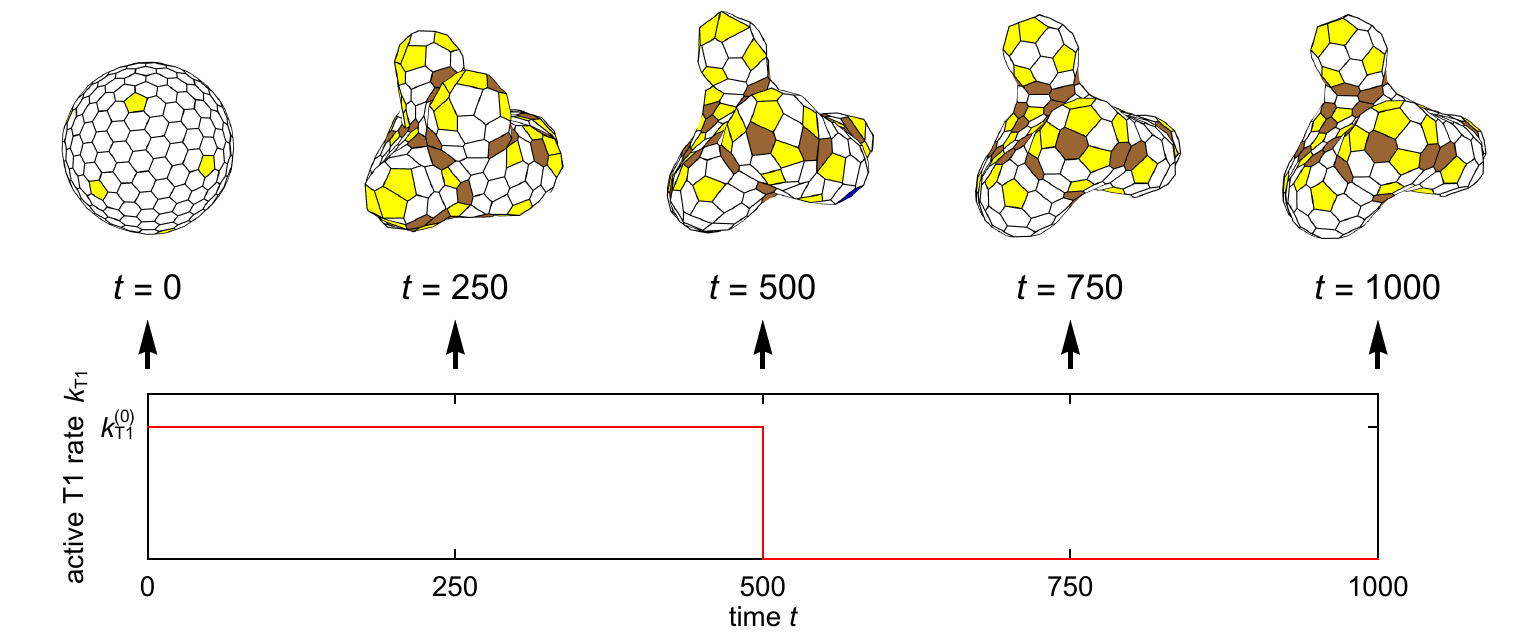}
		\caption{\textbf{Sequence of shapes during step-like active T1 transitions scheme.} Shapes of $\alpha=0.7,\beta=0.5$ model organoids at different stages of the step-like protocol. While the initial $t=0$ shape is spherical, the $t=250$ shape shows that branches are formed rather quickly during the active period. The similarity of the $t=250$ and the $t=500$ shapes suggests that the duration of the active period is not very important as long as it is long enough, and that further extending it would not affect the shapes very much.}
	\end{figure}
	
	\begin{figure}[H]
		\centering
		\includegraphics{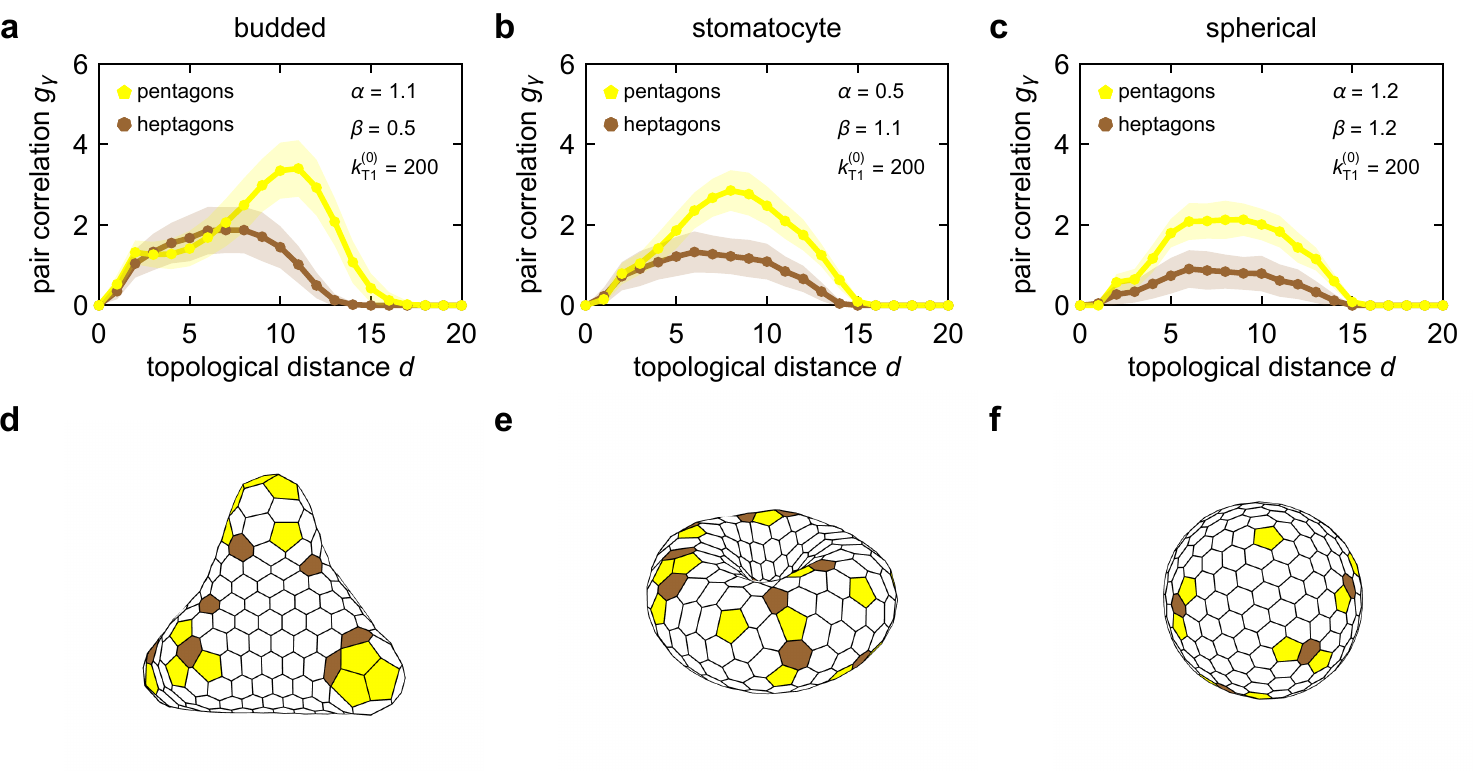}
		\caption{\textbf{Topology-geometry coupling is also seen in budded morphologies, but is absent in stomatocyte and spherical model organoids.} Topological pair correlation function $g_\gamma(d)$ of pentagonal or heptagonal cells in budded, stomatocyte, and spherical shapes ({\bf a}, {\bf b}, and {\bf c}, respectively, with model parameters given in the legend). Solid lines are guides to the eye through averages over 300 instances of shapes shown by points, whereas shaded areas show the standard deviation. The correlation function in the budded shapes is similar to that in the branched shapes (Fig.~2c of the main text), but its features are less pronounced. In the stomatocytes and in the spherical shapes, the form of $g_\gamma$ approaches $\sin d$ expected in a random distribution of points on a sphere, with little or no difference in the topological distance corresponding to the maximal value for pentagons and heptagons. Note that the magnitudes of $g_5$ and $g_7$ are different because the total number of pentagons $n_5$ exceeds the total number of heptagons $n_7$ by $12$ for topological reasons. ({\bf d}-{\bf f})~Representative shapes for the set of model parameters corresponding to each organoid type. }
	\end{figure}
	
	\begin{figure}[H]
		\centering
		\includegraphics{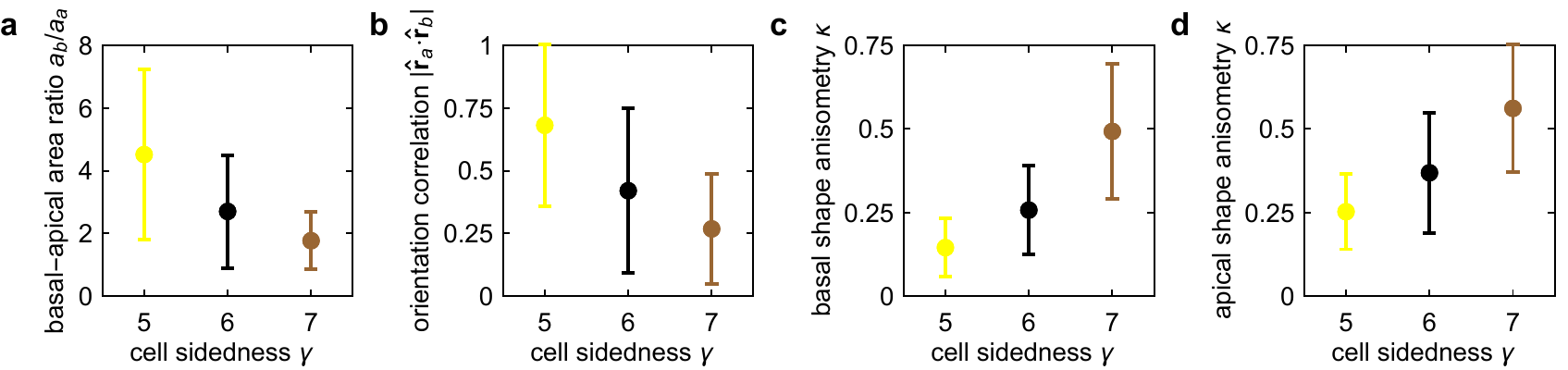}
		\caption{\textbf{Characteristics of cell shape in branched organoids depend significantly of cell sidedness.} {\bf a}~Average ratio of the area of the basal and the apical cell side in pentagonal, hexagonal and heptagonal cells for the model organoid in Fig.~2d of the main text [$\alpha=0.7,\beta=0.5,k_{\rm T1}^{(0)}=200$]. Error bars denote standard deviation. {\bf b}~Average correlation between the long axes of the apical and basal cell sides ($\hat{\mathbf{r}}_a$ and $\hat{\mathbf{r}}_b$, respectively; Methods) for the same model organoid, computed as the magnitude of the dot product of unit vectors pointing along the two long axes. {\bf c} and {\bf d} Shape anisometry $\kappa$ of basal and apical cells, respectively, in the model organoid in Fig.~2d of the main text, with larger values corresponding to more anisometric shapes (Methods).}
	\end{figure}

	\begin{figure}[H]
		\centering
		\includegraphics{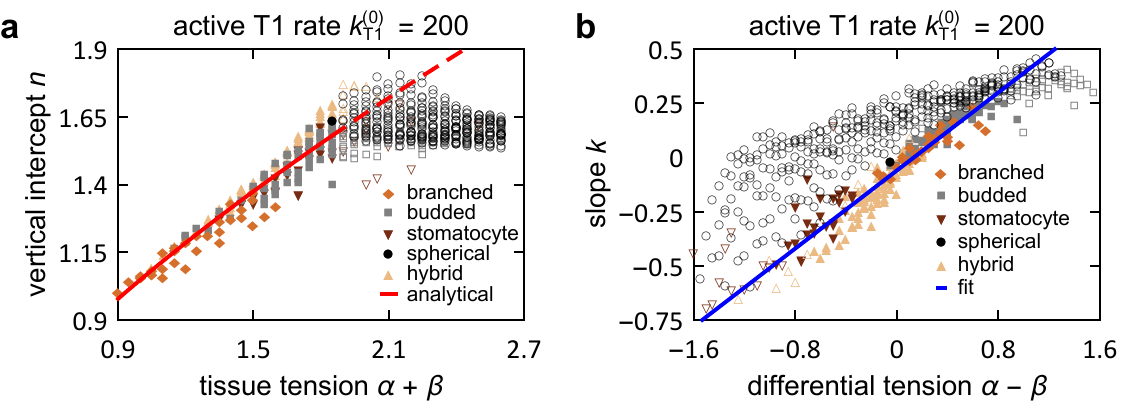}
		\caption{\textbf{Analysis of curvature-thickness coupling.} {\bf a}~Vertical intercepts as a function of tissue tension $\alpha+\beta$ obtained by fitting the $h(c)$ plot by linear functions for all organoids in Fig.~1f of the main text. Points with $\alpha+\beta>1.9$ where the model organoids are spherical are denoted by hollow markers. The red line shows the function $\left (2^{1/3}/3^{1/6}\right )(\alpha+\beta)^{2/3}$, its dashed section corresponding to $\alpha+\beta>1.9$. {\bf b}~Slopes $k$ for the same organoids as a function of differential tension $\alpha-\beta$. Hollow markers are again for points with $\alpha+\beta>1.9$. Blue line shows the linear fit of the type $\lambda_1+(\alpha-\beta)\lambda_2$.}
	\end{figure}
	\newpage\newpage\newpage

\section*{Supplementary Note 1}
\subsection*{\label{ap:cont}3D continuum theory}
	
\begin{figure}[h]
	\centering
	\includegraphics{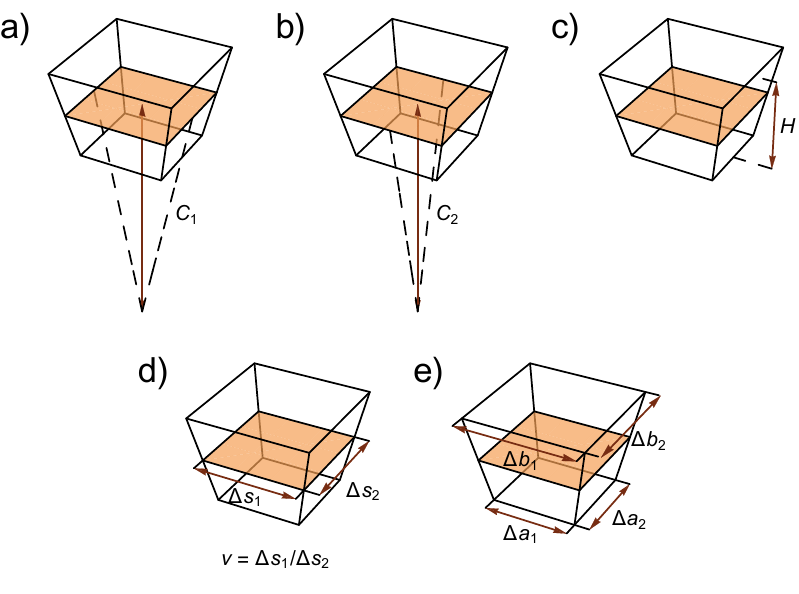}
	\caption{\textbf{Parametrization of model cell.} The model cell represented as a truncated pyramid of volume $V_{\rm cell}$ is defined by the principal curvatures $C_1$ and $C_2$ (a and b, respectively), height $H$ (c), and aspect ratio of the midplane $\nu$ (d). Lengths of edges of the apical and the basal side which depend on $C_1$, $C_2$, $H$, and $\nu$ are shown in panel e.}
	\label{skica}
\end{figure}

The derivation of the continuum theory is based on generalizing the approach reported in Ref.~\cite{Krajnc15} to three dimensions. We begin by defining a model cell represented by a truncated pyramid with a rectangular base and a volume of $V_{\rm cell}$. The model cell is defined by the principal curvatures $C_1$ and $C_2$, height $H$, and the aspect ratio in the midplane located halfway between the apical and the basal side defined by 
\begin{equation}	
    \nu=\frac{\Delta s_1}{ \Delta s_2},
    \label{eq:nu}
\end{equation} 
where $\Delta s_1$ and $\Delta s_2$ are the length and the width of the rectangular midplane cross section, respectively (Supplementary Fig.~\ref{skica}). These four parameters can describe four deformation modes: (i) and (ii) bending in the directions of the principal curvatures, (iii) changes in the local tissue thickness, and (iv) pure shear deformation.

The geometrical parameters of the model cell can be related to the length and the width of the apical side $\Delta a_1$ and $\Delta a_2$, respectively, and the length and the width of the basal sides $\Delta b_1$ and $\Delta b_2$, respectively, by 

\begin{equation}
	\Delta a_1=\frac{1-C_1H/2}{1+C_1H/2}\Delta b_1\label{b1}
\end{equation}
and
\begin{equation}	
	\Delta a_2=\frac{1-C_2H/2}{1+C_2H/2}\Delta b_2.\label{b2}
\end{equation}
Upon inserting the relations $\Delta s_1 = (\Delta a_1 + \Delta b_1) / 2$ and $\Delta s_2 = (\Delta a_2 + \Delta b_2) / 2$ into Eq.~(\ref{eq:nu}) and using Eqs.~(\ref{b1}) and (\ref{b2}), we obtain
\begin{equation}
	\Delta b_2 = \frac{1}{\nu}\frac{1+C_2H/2}{1+C_1H/2}\Delta b_1.\label{a2}
\end{equation}
Lastly, by combining the expression for the midplane area $\Delta A_m =\Delta s_1 \Delta s_2$ with the above relations, we find that
\begin{equation}
	\Delta  b_1^2 =\nu\left(1+C_1H/2\right)^2\Delta A_m.\label{a1}
\end{equation}
Equations~(\ref{b1})-(\ref{a1}) allow us to express the lengths of edges of the apical and basal sides in terms of $C_1$, $C_2$, $\nu$, $H$, and $\Delta A_m$.
	
We now wish to derive the elastic energy per unit area. For a section of tissue with midplane area $\Delta A_m$ at fixed $C_1$, $C_2$, $\nu$, and $L$, the energy reads
\begin{equation}
	\Delta W = \Gamma_a \Delta A_a + \Gamma_b \Delta A_b + \frac{\mathrm{d} W_l}{\mathrm{d} A_m} \Delta A_m\>,\label{eq:energy_start}
\end{equation}
where $\Gamma_a$, $\Gamma_b$, and $\Gamma_l$ are the apical, basal and lateral surface tensions, respectively, $\Delta A_a$ and $\Delta A_b$ are the apical and basal areas corresponding to $\Delta A_m$, whereas $\mathrm{d} W_l / \mathrm{d} A_m$ is the energy density associated with the lateral sides. The energy terms associated with the apical and the basal sides can be rewritten by using $\Delta A_a = \Delta a_1 \Delta a_2$, and $\Delta A_b = \Delta b_1 \Delta b_2$ along with Eqs.~(\ref{b1})-(\ref{a1}) such that
\begin{equation}
	\Gamma_a \Delta A_a + \Gamma_b \Delta A_b = 
		(\Gamma_a + \Gamma_b)\left(1+\frac{C_1C_2H^2}{4}\right)\Delta A_m+(\Gamma_b-\Gamma_a)\frac{(C_1+C_2)H}{2}\Delta A_m.\label{eq:apicobasal_term}
\end{equation}
The derivation of the term associated with the lateral sides is more involved. We first calculate the energy of the lateral sides of a single cell with a volume of $V_{\rm cell}$ and fixed $C_1$, $C_2$, $\nu$, and $H$. This energy reads
\begin{equation}
	{W_l=\frac{1}{2}\Gamma_l(2A_{l1} + 2A_{l2}),\label{Wl}}
\end{equation}
where $A_{l1}$ is the area of the lateral side with edges $\Delta a_1$ and $\Delta b_1$ whereas $A_{l2}$ is the area of the lateral side with edges $\Delta a_2$ and $\Delta b_2$. By using simple trigonometry and relations derived above, $A_{l1}$ and $A_{l2}$ can be recast as
\begin{equation}
	A_{l1}=\frac{1}{1+C_1H/2}\Delta b_1\sqrt{H^2+\left(\frac{C_2H/2}{1+C_1H/2}\right)^2\frac{1}{\nu^2}\Delta b_1^2}
\end{equation}
and
\begin{equation}		
	A_{l2}=\frac{1}{\nu}\frac{1}{1+C_1H/2}\Delta b_1\sqrt{H^2+\left(\frac{C_1H/2}{1+C_1H/2}\right)^2\Delta b_1^2},
\end{equation}
respectively. Now we combine these results with the expression for the volume of a single cell with given $\Delta a_1$, $\Delta a_2$, $\Delta b_1$, $\Delta b_2$, and $H$, which reads
\begin{eqnarray}
	V_{\rm cell} &=& \left(\Delta a_1\Delta a_2 + \frac{1}{2}\Delta a_1\Delta b_2 \nonumber+ \frac{1}{2}\Delta b_1\Delta a_2\Delta b_1\Delta b_2\right)\frac{H}{3}\nonumber\\
	&=&\frac{1+C_1C_2H^2/12}{\left(1+C_1H/2\right)^2}\frac{1}{\nu} \Delta b_1^2 H,
\end{eqnarray}
and we obtain
\begin{equation}
	A_{l1} = \sqrt{\nu}\frac{1}{\sqrt{1+C_1C_2H^2/12}}\sqrt{\frac{V_{\rm cell}}{H}}\sqrt{H^2+\frac{1}{\nu}\frac{\left(C_2H/2\right)^2}{1+C_1C_2H^2/12}\frac{V_{\rm cell}}{H}}
\end{equation}
and			
\begin{equation}
	A_{l2} = \frac{1}{\sqrt{\nu}}\frac{1}{\sqrt{1+C_1C_2H^2/12}}\sqrt{\frac{V_{\rm cell}}{H}}\sqrt{H^2+\nu\frac{\left(C_1H/2\right)^2}{1+C_1C_2H^2/12}\frac{V_{\rm cell}}{H}},
\end{equation}
so that the lateral energy of a single cell reads
%
\begin{eqnarray}
W_l =
	\Gamma_l\frac{1}{\sqrt{1+C_1C_2H^2/12}}\sqrt{\frac{V_{\rm cell}}{H}}\left[\sqrt{\nu}\sqrt{H^2+\frac{1}{\nu}\frac{\left(C_2H/2\right)^2}{1+C_1C_2H^2/12}\frac{V_{\rm cell}}{H}} +  \frac{1}{\sqrt{\nu}}\sqrt{H^2+\nu\frac{\left(C_1H/2\right)^2}{1+C_1C_2H^2/12}\frac{V_{\rm cell}}{H}} \right].
\end{eqnarray}
%
By comparing this expression with $\left(\mathrm{d} W_l / \mathrm{d} A_m\right) \Delta A_m$ and expressing $\Delta A_m$ of the cross-section of the single cell as 
\begin{eqnarray}
	\Delta A_m = \frac{1}{1+C_1C_2H^2/12}\frac{V_{\rm cell}}{H},
\end{eqnarray}
we find that
\begin{eqnarray}
	\frac{\mathrm{d} W_l}{\mathrm{d} A_m} &= &\Gamma_l\sqrt{1 + \frac{1}{12}C_1C_2H^2}\sqrt{\frac{H^3}{V_{\rm cell}}}\nonumber\\ & &  \times\left[\sqrt{\nu}\sqrt{1+\frac{1}{\nu}\frac{1}{1+C_1C_2H^2/12}\frac{C_2^2V_{\rm cell}}{4H}}+\frac{1}{\sqrt{\nu}}\sqrt{1+\nu\frac{1}{1+C_1C_2H^2/12}\frac{C_1^2V_{\rm cell}}{4H}}\right].
\end{eqnarray}
Upon inserting this result along with Eq.~(\ref{eq:apicobasal_term}) into Eq.~(\ref{eq:energy_start}), we obtain the final energy density of a tissue section of midplane area $\Delta A_m$. To recast it in dimensionless form, we divide it by $\Gamma_l V_{\rm cell}^{2/3}$ and introduce the dimensionless energy $w = W / \big(\Gamma_l V_{\rm cell}^{2/3}\bigr)$, dimensionless midplane area $A = A_m / V_{\rm cell}^{2/3}$, dimensionless cell height $h = H / V_{\rm cell}^{1/3}$, dimensionless curvatures $c_1 = C_1 V_{\rm cell}^{1/3}$ and $c_2 = C_2 V_{\rm cell}^{1/3}$, dimensionless apical tension $\alpha = \Gamma_a / \Gamma_l$, and dimensionless basal tension $\beta = \Gamma_b / \Gamma_l$. Upon taking the limit $\Delta A_m \rightarrow 0$, we find that the final expression for the dimensionless energy per unit dimensionless area is
%
\begin{eqnarray}
	\frac{\mathrm{d} w}{\mathrm{d}A } &=& (\alpha + \beta) \left(1+\frac{c_1c_2h^2}{4}\right)+(\beta-\alpha)\frac{(c_1+c_2)h}{2}\nonumber \\
	& & +\sqrt{1+\frac{1}{12}c_1c_2h^2}\sqrt{h^3}\left[\sqrt{\nu}\sqrt{1+\frac{1}{\nu}\frac{1}{1+c_1c_2h^2/12}\frac{c_2^2}{4h}}+\frac{1}{\sqrt{\nu}}\sqrt{1+\nu\frac{1}{1+c_1c_2h^2/12}\frac{c_1^2}{4h}}\right].
\end{eqnarray}
%
{
This result is rather involved but it can be considerably simplified if $c_1^2/4h$, $c_2^2/4h$, and $c_1c_2h^2/12$ are all considerably smaller than unity. (For example, the mean values of $c^2/4h$ and $c^2h^2/12$ in the organoids in Figs.~3a and b of the main text are 0.026 and 0.025, respectively.) In this limit we find that
\begin{eqnarray}
	\frac{\mathrm{d} w}{\mathrm{d}A } &=& (\alpha + \beta) \left(1+\frac{c_1c_2h^2}{4}\right)+(\beta-\alpha)\frac{(c_1+c_2)h}{2}\label{secondOrder}\nonumber\\& &+ \left(\sqrt{\nu}+\frac{1}{\sqrt{\nu}}\right)h^{3/2}+\frac{1}{24}\left(\sqrt{\nu}+\frac{1}{\sqrt{\nu}}\right)c_1c_2h^{7/2}+\frac{1}{8}h^{1/2}\left(\sqrt{\nu}c_1^2+\frac{1}{\sqrt{\nu}}c_2^2\right).
\end{eqnarray}
Note that the form of Eq.~(\ref{secondOrder}) is also obtained in the limit of small deformations where $c_1h \rightarrow 0$ and $c_2h \rightarrow 0$; this limit is however less relevant for our organoid shapes.}

We now further assume that the tissue is fluidized so that $\nu=1$, and we rearrange the above expression as
\begin{eqnarray}
	\frac{\mathrm{d} w}{\mathrm{d}A } &= &\underbrace{(\alpha+\beta)+2\left[1-\frac{(\alpha-\beta)^2}{4}\right]\sqrt{h^3}}_{\rm{surface\ tension}}\label{secondOrderFluidized}\nonumber\\
	& &+\underbrace{
	\frac{\sqrt{h}}{8}\left[c_1+c_2-2\sqrt{h}(\alpha-\beta)\right]^2
	}_\mathrm{local\ bending\ energy}\nonumber\\
	& &+\underbrace{
	\left(\frac{\alpha+\beta}{4}+\frac{\sqrt{h^3}}{12}-\frac{1}{4\sqrt{h^3}}\right)h^2c_1c_2
	}_\mathrm{Gaussian\ bending\ energy}.
\end{eqnarray}
The first term, which is independent of curvature, represents the surface tension of the tissue, the second one is the local bending energy with a non-zero spontaneous curvature  analogous to the bending energy of lipid membranes within the spontaneous-curvature model \cite{Seifert91}, whereas the third term is the Gaussian bending energy. The analogy with lipid membranes is not complete as all three terms also depend on the local tissue thickness $h$; in lipid membranes, thickness is assumed to be constant. 
	
To estimate the local bending modulus $k_c$, the Gaussian modulus $k_G$, and the spontaneous curvature $c_0$, we replace the actual value of $h$ (which is a variable rather than a constant) by the equilibrium cell height corresponding to a flat epithelium, which reads $h = (\alpha+\beta)^{2/3}$; here we assumed that the apical and the basal sides are squares. This gives
\begin{equation}
	k_c = \frac{1}{4}(\alpha+\beta)^{1/3},
\end{equation}
\begin{equation}
	k_G = \left[\frac{1}{3}(\alpha+\beta)^2-\frac{1}{4}\right](\alpha+\beta)^{1/3},
\end{equation}
and
\begin{equation}
	c_0=2(\alpha+\beta)^{1/3}(\alpha-\beta)\label{c0}
\end{equation}
The curvature-thickness coupling term $(\beta-\alpha)(c_1+c_2)h/2$ in Eq.~(\ref{secondOrder}), which gives rise to the spontaneous curvature in Eq.~(\ref{secondOrderFluidized}), can also be interpreted in a different manner. We again consider a fluidized tissue with $\nu=1$ and we assume that the tissue forms a closed shell as in the main text. We then integrate the energy per unit area over the entire surface, which gives
\begin{equation}
	w=\oint_A\left[\left(\alpha + \beta\right) \left(1+\frac{c_1c_2h^2}{4}\right)+2h^{3/2}+ \frac{1}{12}c_1c_2h^{7/2}+\frac{1}{8}h^{1/2}\left(c_1^2+c_2^2\right)\right]\mathrm{d}A +\frac{\beta-\alpha}{2}\oint_A(c_1+c_2)h\, \mathrm{d}A.
\end{equation}
If $h$ does not vary across the shell, then the integral $\oint_{A}(c_1+c_2)h\, \mathrm{d}A$ is equal to the difference between the apical and the basal area of the shell $\Delta A$, and we have
\begin{eqnarray}
	w&=f(\alpha+\beta)+\frac{1}{2}(\beta-\alpha)\Delta A,
\end{eqnarray}
where $f$ is a shape-dependent function of total tension $\alpha+\beta$ but not of differential tension $\beta-\alpha$. From the form of this expression we see that the differential tension $\beta-\alpha$ is coupled to $\Delta A$ and thus $(\beta-\alpha)/2$ effectively acts as an area-difference tension. In this respect, the continuum theory of our model tissues derived above departs from the area-difference-elasticity theory of lipid vesicles where the so-called non-local bending energy gives rise to a term $\propto(\Delta A-\Delta A_0)^2$~\cite{Svetina89}.

{

As an estimate of the error arising from rectangular rather than a hexagonal base, our continuum theory gives the height of a flat epithelium as $(\alpha+\beta)^{2/3}$, whereas the exact result for a flat epithelium of hexagonal cells is $\left (2^{1/3}/3^{1/6}\right )(\alpha+\beta)^{2/3}$, a difference of only about $5\%$; presumably, a similar relatively small scale of error can be assumed for $c_1$ and $c_2$, and all  terms are higher than first order. 
}

\section*{Supplementary Movies}
	{ 
	\noindent Movie 1. Evolution of shapes of the three model organoids in Fig.~1c of the main text without active T1 transitions ($k_{\rm T1}^{(0)} = 0$), $N_c = 300$, and $v_{\rm lumen} = 100$. The simulations start from a spherical shape and result in a spherical ($\alpha = 1.2,\>\beta = 1.2$), stomatocyte ($\alpha = 0.5,\>\beta = 1.1$), and budded ($\alpha = 1.5,\>\beta = 0.3$) morphology.\\ 
	}
	
	\noindent Movie 2. Evolution of shapes of the four active T1 model organoids in Fig.~1d of the main text, with $k_{\rm T1}^{(0)} = 200$, $N_c = 300$, and $v_{\rm lumen} = 100$. The simulations start from a spherical shape and result in a spherical ($\alpha = 1.2,\>\beta = 1.2$),  stomatocyte ($\alpha = 0.5,\>\beta = 1.1$), budded ($\alpha = 1.1,\>\beta = 0.5$),  and branched ($\alpha = 0.7,\>\beta = 0.5$) morphology.\\ 

    \noindent Movie 3. Evolution of the branched model organoid in Fig.~2d of the main text ($\alpha = 0.7,\>\beta = 0.5$). Cells with five neighbors are colored yellow, cells with seven neighbors are colored brown, and all other cells are white.\\ 
    
    \noindent Movie 4. Evolution of shapes of the four growing model organoids with $\tau_D = 2000$ in Figs.~4b-e of the main text. Note that while in the  spherical ($\alpha=1.2,\>\beta=1.2$), stomatocyte ($\alpha=0.5,\>\beta=1.1$), and budded ($\alpha=1.1,\>\beta=0.5$) organoid the shapes resulting from growth and from junctional activity (Fig.~1d of the main text) belong to the same category, the ($\alpha=0.7,\>\beta=0.5$) growth-induced shape differs from the branched morphology in Fig.~1d as branching requires the presence of active T1 transitions, which are absent in the growing-organoid model.\\
    
    {
    \noindent Movie 5. Evolution of shapes of the two growing model organoids with $\tau_D = 2000$ in Figs.~4f-g of the main text with ($\alpha=0.7,\>\beta=0.5$), without active T1 transitions ($k_{\rm T1}^{(0)} = 0$; left), and with active T1 transitions ($k_{\rm T1}^{(0)} = 200$; right).
    }